\documentclass[onecolumn]{aa}
\usepackage{graphicx}

\begin{document}

\title{Solar wind and motion of dust grains}
\author{J. Kla\v{c}ka, J. Petr\v{z}ala, P. P\'{a}stor, L. K\'{o}mar}
\institute{Faculty of Mathematics,
Physics and Informatics, Comenius University \\
Mlynsk\'{a} dolina, 842 48 Bratislava, Slovak Republic \\
e-mail: klacka@fmph.uniba.sk}

\date{}

\abstract{
Action of solar wind on arbitrarily shaped interplanetary dust particle is
investigated. The final relativistically covariant equation of motion of
the particle contains both orbital evolution and change of particle's mass.
Non-radial solar wind velocity vector is also included.
The covariant equation of motion reduces
to the Poynting-Robertson effect in the limiting case when spherical
particle is treated, the speed of the
incident solar wind corpuscles tends to the speed of light and 
the corpuscles spread radially from the Sun. The results of quantum 
mechanics have to be incorporated into the physical considerations,
in order to obtain the limiting case.
The condition for the solar wind effect on motion of spherical
interplanetary dust particle is $\vec{p}'_{out}$ $=$
( 1 $-$ $\sigma'_{pr} / \sigma'_{tot}$ ) $\vec{p}'_{in}$, where
$\vec{p}'_{in}$ and $\vec{p}'_{out}$ are incoming and outgoing radiation
momenta (per unit time) measured in the proper frame of reference of the
particle; $\sigma'_{pr}$ and $\sigma'_{tot}$ are solar wind pressure and
total scattering cross sections.

Analytical solution of the derived equation of motion yields
qualitative behaviour consistent with numerical calculation.
This holds also if decrease of particle's mass is important,
when also outspiralling from the Sun may occur.

Real flux density of radial solar wind energy yields that 
the time of spiralling toward the Sun may differ in about 10 \% 
from the case of constant flux for heliocentric distances 
smaller than about 10 AU (if radius of the particle is unchaged). 
The differences between the real solar wind effect 
(when also non-radial solar wind velocity vector is taken into account)
and the standard approach may be even more significant for heliocentric
distances greater than 10 AU: solar wind can cause outspiralling from 
the Sun, even if radius of the particle is constant.

Real flux density of solar wind energy produces shift of perihelion
of interplanetary dust particles. This result significantly
differs from the standard treatment of the action of the solar wind 
on dust particles, when analogy with the Poynting-Robertson effect is 
stressed. Moreover, the evolution of the shift of perihelion depends 
on orbital position of the parent body at the time of ejection of the 
particle.

}

\authorrunning{Kla\v{c}ka et al.}
\titlerunning{Solar wind and motion of dust grain}

\maketitle

\section{Introduction}
The Poynting-Robertson effect is used in modelling of
orbital evolution of dust grains under the action of electromagnetic
radiation (of the central star), for many decades (e. g., Poynting 1903,
Robertson 1937, Wyatt and Whipple 1950, Dohnanyi 1978,
Kapi\v{s}insk\'{y} 1984, Jackson and Zook 1989,
Leinert and Gr\H{u}n 1990, Gustafson 1994,  Dermott {\it et al.} 1994,
Reach {\it et al.} 1995). It was presented that solar wind
operates in a similar way.
The action of the solar wind on motion of interplanetary dust particle
was discussed, in a heuristic way, e. g., by Whipple (1955).
He has mentioned also the
results of laboratory experiments: intense bombardment of a material
by energetic corpuscles destructs the material and this effect
is known as a "sputtering". Current opinion is that we have two different
effects of the solar wind: i) motion of dust particle is influenced by
the incident solar wind, and, ii) the corpuscular sputtering  
decreases mass of the particle (e. g., Whipple 1955, Dohnanyi 1978,
Kapi\v{s}insk\'{y} 1984, Leinert and Gr\"{u}n 1990). There has
been an attempt to better understand physics of the action of the
solar wind on the motion of dust particle. As a first attempt,
we can mention Robertson and Noonan (1968, pp. 122-123).
The authors formulated relativistically covariant
equation of motion of the particle under the action of the solar wind.
However, their result does not admit any destruction of the particle.
A more realistic view was presented in Kla\v{c}ka and Saniga (1993),
where also space-time formulation of the problem was suggested. As a
result, corpuscular sputtering is an indispensable part of the equation
of motion for the action of solar wind on interplanetary dust particle.

Our paper presents space-time formulation of the action of solar wind
on arbitrarily shaped interplanetary dust particle.
Equation of motion in a relativistically covariant form is derived.
Moreover, in order to be physically correct, results of quantum theory
are also taken into account. The results of the paper are consistent
with the results of the papers by Kla\v{c}ka (2008a, 2008b)
for the electromagnetic radiation. Our theoretical derivations hold
for any solar wind velocity vector and the result can be easily applied
to other stars with stellar winds.

Application of the derived equation of motion to spherical 
interplanetary dust particle is presented in the form of orbital 
evolution of the particle. While submicron dust particles are driven 
mainly by Lorentz force (motion of charged particles in the 
interplanetary magnetic field, Dohnanyi 1978, Leinert and Gr\"{u}n 1990, 
Dermott {\it et al.} 2001), collisions among particles are important 
for particles of radii larger
than hundred of micrometres, approximately (Gr\"{u}n {\it et al.} 1985,
Dermott {\it et al.} 2001). This paper deals with orbital evolution of
micron-sized spherical interplanetary dust particles,
when effects of solar gravity,
solar electromagnetic radiation and solar wind (solar corpuscular
radiation) are relevant. Radial solar wind is conventionally used.
However, the newest observations ( Bruno {\it et al.} 2003) show
that velocity vector of the solar wind corpuscles is non-radial
and the angle between the velocity vector and radial direction
is practically independent on heliocentric distance.
Our paper compares orbital evolution of spherical interplanetary dust particle
for standardly used approach of time independent radial solar wind
and more real solar wind model. Also the action of mass decrease of the 
interplanetary dust particle is taken into account.

Section 2 derives relativistically covariant equation of motion
of an arbitrarily shaped particle under the action of solar wind
(including non-radial component of the solar wind velocity).
Sec. 3 summarizes important equations for the Poynting-Robertson effect.
Equation of motion of the spherical particle under the action of
solar corpuscular and electromagnetic radiation, and, solar gravity,
is given in Sec. 4.
Sec. 5 deals with secular evolution of particle's orbital elements under the
action of solar radiation (electromagnetic and corpuscular -- solar wind),
in an analytical way.
The following section Sec. 6 concentrates on detail treatment of the numerical
results and compares the results for conventional time independent 
radial solar wind with those obtained for more real solar wind model.

\section{Equation of motion - solar wind effect}
Relativistically covariant equation of motion of interplanetary dust particle
is derived in this section. Our derivation enables understand physics
of the action of the solar wind on the motion of dust particle (compare
heuristic explanation by Whipple 1955 and space-time formulations
presented by  Robertson and Noonan 1968, pp. 122-123, and,
Kla\v{c}ka and Saniga 1993).

We will show that the corpuscular sputtering is an indispensable 
part of the action of the solar wind on the interplanetary dust particle, 
so the sputtering cannot be considered as an another effect of the 
solar wind. Moreover, non-radial solar wind velocity vector can be easily incorporated into the final equation of motion. Finally, the covariant formulation yields the Poynting-Robertson effect in the limiting case 
when speed of the solar wind corpuscles tends to the speed of light. 
The limiting case is fulfilled under the assumption that the total 
cross section of the interaction between the solar wind corpuscles
and the interplanetary dust particle is given by the results of 
quantum theory and not by the classical non-quantum physics.

This section presents also results with an accuracy to the order
$(\vec{v}/u)^{2}$, where $\vec{v}$ is orbital velocity of the particle
with respect to the Sun and $u$ is the solar wind speed. These results
will be used in practical modelling of orbital evolution of interplanetary
dust grains in Secs. 4 and 5.

\subsection{Incident radiation}
Let us introduce two inertial reference frames. The first is the proper
reference frame of a particle moving with velocity $\vec{v}$
around the Sun. The particle is at rest in it's proper frame of reference.
Quantities measured in this frame will be primed. The second frame is
associated with the Sun. This frame is "stationary reference frame".

We will suppose that all corpuscles of the solar wind are of the same
mass $m_{1}$ and of the same velocity $\vec{u}$ (or $\vec{u'}$ in the
proper reference frame of the particle). Thus, each of the corpuscles
has the following four-momentum
\begin{equation}\label{1}
{p'}_{1}^{\mu} = \left(E'_{1}/c ~;~\vec{p'}_{1}\right) =
m_{1} \gamma \left(u'\right) \left(c~;~\vec{u'}\right)
\end{equation}
in the proper reference frame of the interplanetary dust particle, or
\begin{equation}\label{2}
p_{1}^{\mu}~=~\left( E_{1}/c~;~\vec{p}_{1}\right)~=~
m_{1} \gamma \left ( u \right ) \left ( c~;~\vec{u} \right )
\end{equation}
in the stationary reference frame; $c$ is the speed of light.

Let a beam of such solar wind corpuscles hits the dust grain.
Energy  and momentum incident on the particle
per unit time in its proper frame are
\begin{eqnarray}\label{3}
E'_{in} &=& \sigma'_{tot} ~n' ~u' ~E'_{1}~,
\nonumber\\
\vec{p'}_{in} &=& \sigma'_{tot} ~n' ~u' ~\vec{p'}_{1}~,
\end{eqnarray}
where $n'$ is the concentration of the solar wind corpuscles and
$\sigma'_{tot}$ is the total scattering cross section of the interplanetary dust
particle. Using Eq. (1), we can rewrite Eqs. (3) into the form of
the incident four-momentum per unit time
\begin{eqnarray}\label{4}
{p'}_{in}^{\mu} &=& \sigma'_{tot} ~n' ~u'
		 ~\left ( \frac{E'_{1}}{c}~;~\vec{p'}_{1} \right )
\nonumber \\
	       &=& \frac{1}{c}~\sigma'_{tot} ~n' ~u' ~E'_{1}
		  \left(1~;~\frac{\vec{u'}}{c} \right)~.
\end{eqnarray}
Introducing the flux density of the incident energy
(energy flow per unit area perpendicular to the beam
of solar wind corpuscles per unit time)
\begin{equation}\label{5}
S' = n' u' E'_{1} ~,
\end{equation}
Eq. (4) can be rewritten
to the form
\begin{eqnarray}\label{6}
{p'}_{in}^{\mu} &=& \left ( \frac{E'_{in}}{c}~;~\vec{p'}_{in} \right ) ~,
\nonumber \\
{p'}_{in}^{\mu} &=& \frac{1}{c}~S' ~\sigma'_{tot} ~
		 \left ( 1~;~\frac{\vec{u'}}{c} \right ) ~.
\end{eqnarray}

Having a four-vector $B'^{\mu}$ $=$ $(B'^{0}~;~\vec{B'})$ in the
proper reference frame, the components of the four-vector in the
stationary reference frame are given by generalized special Lorentz transformation:
\begin{eqnarray}\label{7}
B^{0}&=&\gamma\left(v\right)\left(B'^{0}~+~\frac{\vec{v}\cdot\vec{B'}}
{c}\right)~,
\nonumber \\
\vec{B}&=& \vec{B'}~+~\left \{ \left [ \gamma \left( v \right) - 1 \right ]
\frac{\vec{v}\cdot\vec{B'}}{v^{2}}~+~\frac{\gamma\left( v \right)}{c}
~B'^{0}\right\} \vec{v}~,
\end{eqnarray}
or inverse

\begin{eqnarray}\label{8}
B'^{0} &=& \gamma\left(v\right)\left(B^{0}~-~\frac{\vec{v}\cdot\vec{B}}
{c}\right)~,
\nonumber\\
\vec{B'} &=& \vec{B}~+~\left \{ \left [ \gamma\left(v\right) - 1 \right]
\frac{\vec{v}\cdot\vec{B}}{v^{2}}~-~\frac{\gamma\left(v\right)}{c}
~B^{0}\right \} \vec{v}~.
\end{eqnarray}

Now, using Eqs. (6) and (7) we get
\begin{eqnarray}\label{9}
p_{in}^{0} &=& \frac{1}{c}~S' ~\sigma'_{tot} ~ \gamma \left ( v \right )
\left ( 1 ~+~ \frac{\vec{v}\cdot\vec{u'}}{c^{2}} \right ) ~,
\nonumber \\
\vec{p}_{in} &=& \frac{1}{c^{2}}~S' ~\sigma'_{tot} ~ \left\{\vec{u'}+\left[
\left(\gamma\left(v\right) -1 \right ) ~\frac{\vec{v} \cdot \vec{u'}}{v^{2}}
~+~ \gamma \left ( v \right ) \right ] \vec{v} \right \} ~.
\end{eqnarray}
We have to express the primed quantities (except of $\sigma'_{tot}$) on the
right-hand sides of Eqs. (9), i.e. $S'$ $=$ $n' u' E'_{1}$ and $\vec{u'}$,
through the unprimed quantities measured in the stationary reference frame of
the Sun. The energy $E'_{1}$ we obtain from Lorentz transformation of
$p_{1}^{\mu}$ to the proper reference frame of the interplanetary dust
particle. It holds
\begin{equation}\label{10}
E'_{1}~=~\gamma\left(v\right) \left(E_{1}~-~\vec{v}\cdot\vec{p}_{1}
\right)~=~ \gamma\left(v\right) \left(1~-~\frac{\vec{v}\cdot
\vec{u}}{c^{2}}\right) ~E_{1} =~\omega~E_{1} ~,
\end{equation}
where we defined the quantity
\begin{equation}\label{11}
\omega~\equiv~\gamma\left(v\right) \left(1~-~\frac{\vec{v}\cdot
\vec{u}}{c^{2}}\right)~.
\end{equation}

Other quantities we get from transformation of the four-vector
of the current density $j^{\mu}$ $=$ $(n c~;~n \vec{u})$ to the
corresponding four-vector ${j'}^{\mu}$ $=$ $(n' c~;~n' \vec{u'})$. The transformation yields
\begin{eqnarray}\label{12}
n' &=& \omega ~n ~,
\nonumber\\
\vec{u'} &=& \frac{1}{\omega}~\vec{\alpha}~,
\end{eqnarray}
where the vector
\begin{equation}\label{13}
\vec{\alpha}~\equiv~\vec{u}+\left[\left(\gamma\left(
v\right ) - 1 \right ) ~\frac{\vec{v}\cdot\vec{u}}{v^{2}}~-~\gamma \left( v \right)
\right] \vec{v} ~,
\end{equation}
has magnitude
\begin{equation}\label{14}
\alpha~=~\left \{ u^{2} + \gamma^{2}\left(v\right)v^{2} - 2 \gamma^{2}
\left ( v \right) \vec{v}\cdot\vec{u} + \gamma^{2}\left(v\right)
\left ( \frac{\vec{v}\cdot\vec{u}}{c}\right)^{2}\right \}^{1/2}~.
\end{equation}
Thus, $u'$ $=$ $\alpha / \omega$ and the flux density of energy is
\begin{eqnarray}\label{15}
S' &=& \frac{\alpha~\omega}{u} ~S ~,
\nonumber \\
S &\equiv& n u E_{1} ~,
\end{eqnarray}
according to Eqs. (5), (10) and (12).

Finally, using Eqs. (9), (11), (12), (13) and (15), one obtains
\begin{eqnarray}\label{16}
p_{in}^{0} &=& \frac{1}{c}~\sigma'_{tot} ~S~\frac{\alpha~\omega}{u}
	       ~\frac{1}{\omega} ~,
\nonumber\\
\vec{p}_{in} &=& \frac{1}{c}~\sigma'_{tot} ~S~ \frac{\alpha~\omega}{u}~
\frac{1}{\omega}~\frac{\vec{u}}{c}~.
\end{eqnarray}

The incident four-momentum of solar wind per unit time is
\begin{eqnarray}\label{17}
p_{in}^{\mu}&=&\frac{1}{c}~\sigma'_{tot} ~S~\frac{\alpha~\omega}{u}~\xi^{\mu}~,
\nonumber\\
\xi^{\mu} &\equiv&
\left(\frac{1}{\omega}~;~\frac{1}{\omega}~\frac{\vec{u}}{c}\right)~.
\end{eqnarray}

\subsection{The reaction of spherical dust particle on the incident solar wind}
The incident solar wind corpuscule may be reflected from the surface of
the interplanetary dust particle (IDP) or may cause its
erosion/destruction and decrease the mass of the IDP, or, in general,
similar processes as reflection, absorption and diffraction may occur.
Let the particle's loss of energy (per unit time) in the proper reference
frame of the IDP, be $E'_{out}$ $-$ $E'_{in}$. $E'_{out}$ can be written
as an $x'$-part of the incident energy per unit time. The relation
\begin{equation}\label{18}
E'_{out}~=~x' ~E'_{in}
\end{equation}
holds for the outgoing energy. In order to express the outgoing momentum 
we declare the ortonormal vector basis $\{\vec{f'}_{j}; j=1, 2, 3\}$ 
in the proper reference frame of the particle and 
three velocity vectors $\{\vec{u'}_{j}$ $=$ 
$u'~\vec{f'}_{j};$ $j=1, 2, 3\}$ corresponding to these unit 
vectors. We suppose that $\vec{u'}_{1}$ $\equiv$ $\vec{u'}$. 
Now, the outgoing momentum per unit time is 
\begin{equation}\label{19}
\vec{p'}_{out}~=~\left ( 1 ~-~ \frac{\sigma'_{pr}}{\sigma'_{tot}} \right ) ~
		 \vec{p'}_{in} ~-~ \sigma'_{tot} ~\frac{S'}{c} ~
		 \sum_{j=2}^{3} \frac{\sigma '_{pr, j}}{\sigma '_{tot}}~ 
\frac{\vec{u'}_{j}}{c}~,
\end{equation}
where $\sigma'_{pr, j}$ (j $=$ 1, 2, 3; $\sigma'_{pr, 1}$ $\equiv$
$\sigma'_{pr}$) are pressure cross sections (analogy with optics --
see  Kla\v{c}ka 2008a, 2008b).

The outgoing four-momentum per unit time is given by Eqs. (6), and (18)-(19):
\begin{eqnarray}\label{20}
{p'}_{out}^{\mu} &=& \left ( \frac{E'_{out}}{c}~;~\vec{p'}_{out} \right ) ~,
\nonumber \\
{p'}_{out}^{\mu} &=& \left \{ \frac{1}{c} ~\sigma'_{tot} ~S' ~x'~;~
	\left ( 1 ~-~ \frac{\sigma'_{pr}}{\sigma'_{tot}} \right ) ~
		 \vec{p'}_{in} ~-~ \sigma'_{tot} ~\frac{S'}{c} ~
		 \sum_{j=2}^{3} \frac{\sigma'_{pr, j}}{\sigma'_{tot}}~ 
\frac{\vec{u'}_{j}}{c} \right\}~.
\end{eqnarray}
Generalized special Lorentz transformation of ${p'}_{out}^{\mu}$, using
Eqs. (15), gives for the outgoing four-momentum per unit time in
the stationary reference frame
\begin{eqnarray}\label{21}
p_{out}^{\mu} &=& \frac{1}{c}~\sigma'_{tot} ~S ~\frac{\alpha~\omega}{u} ~x'~
	       \frac{U^{\mu}}{c}
\nonumber \\
& & +~	\left ( 1 ~-~ \frac{\sigma'_{pr}}{\sigma'_{tot}} \right ) ~
	\frac{1}{c}~\sigma'_{tot} ~S ~\frac{\alpha~\omega}{u} ~
	\left ( \xi^{\mu} ~-~ \frac{U^{\mu}}{c}  \right )
\nonumber \\
& & -~	\sigma'_{tot} ~\frac{S}{c} ~ \frac{\alpha ~ \omega}{u} ~\sum_{j=2}^{3}
	\frac{\sigma'_{pr, j}}{\sigma'_{tot}}~ \left ( \xi_{j}^{\mu} ~-~
	\frac{U^{\mu}}{c} \right )~,
\end{eqnarray}
where
\begin{equation}\label{22}
U^{\mu} = \left( \gamma(v)~c~;~ \gamma(v) ~\vec{v}\right)
\end{equation}
is four-velocity of the IDP. The other four-vectors are
\begin{eqnarray}\label{23}
\xi_{j}^{\mu} &=& \left ( \frac{1}{\omega_{j}}~;~ \frac{1}{\omega_{j}}~
		  \frac{\vec{u}_{j}}{c} \right ) ~,
\nonumber \\
\omega_{j} &\equiv& \gamma (v) \left ( 1 ~-~ 
\frac{\vec{v} \cdot \vec{u}_{j}}{c^{2}} \right ) ~,
\nonumber\\
\vec{u}_{j} &=& \left[\gamma \left(v\right)\left(1+
\frac{\vec{v}\cdot\vec{u'}_{j}}{c^{2}}\right)\right]^{-1} \left\{
\vec{u'}_{j}~+~\left[\left(\gamma \left(v\right)-1\right)~
\frac{\vec{v}\cdot\vec{u'}_{j}}{v^{2}}~+~\gamma \left(v\right)\right]
\vec{v} \right\}~,
\nonumber\\
&& j=1, 2, 3~,
\end{eqnarray}
and  $\omega_{1} \equiv \omega$, $\xi_{1}^{\mu}$ $\equiv$ $\xi^{\mu}$, 
$\vec{u}_{1} \equiv \vec{u}$.

\subsection{Equation of motion}
Now we can write equation of motion of the IDP under the action
of the solar wind, in a relativistically covariant form:
\begin{equation}\label{24}
\frac{dp^{\mu}}{d\tau}~=~p_{in}^{\mu}~-~p_{out}^{\mu}~.
\end{equation}
Eq. (24) yields, using Eqs. (17) and (21),
\begin{eqnarray}\label{25}
\frac{dp^{\mu}}{d\tau} &=& \frac{1}{c}~ \sigma'_{tot} ~S~\frac{\alpha~\omega}{u}~
\nonumber \\
& & \times \left \{ \frac{\sigma'_{pr}}{\sigma'_{tot}} ~ \xi^{\mu} ~-~
\left [ x' ~-~  \left ( 1 ~-~ \frac{\sigma'_{pr}}{\sigma'_{tot}} \right ) \right ]
~\frac{U^{\mu}}{c} \right \}
\nonumber \\
& & +~	\sigma'_{tot} ~\frac{S}{c} ~ \frac{\alpha ~ \omega}{u} ~\sum_{j=2}^{3}
	\frac{\sigma'_{pr, j}}{\sigma'_{tot}}~ \left ( \xi_{j}^{\mu} ~-~
	\frac{U^{\mu}}{c} \right)~,
\end{eqnarray}
where $p^{\mu}$ $=$ $m~ U^{\mu}$ is four-momentum of the IDP of mass
$m$ and $\tau$ is the proper time of the particle.

Using
\begin{equation}\label{26}
\frac{dp^{\mu}}{d\tau}~=~\frac{d}{d\tau}~\left ( m ~U^{\mu} \right )~=~
\frac{dm}{d\tau}~U^{\mu}~+~m~\frac{dU^{\mu}}{d\tau}~,
\end{equation}
Eq. (25) yields not only acceleration of the particle, but also
change of the particle's (rest) mass, due to the interaction of
the IDP with the solar wind. The change of the mass is given,
on the basis of Eqs. (11), (17), (22), (25) and (26), by the expression
($U_{\mu} U^{\mu}$ $=$ $c^{2}$, $U_{\mu}$ $dU^{\mu} / d\tau$ $=$ 0)
\begin{equation}\label{27}
\frac{dm}{d\tau}~=~-~ \frac{1}{c^{2}}~ \sigma'_{tot} ~
		   S~\frac{\alpha~\omega}{u} ~\left ( x' ~-~ 1 \right ) ~.
\end{equation}
One can easily verify that Eq. (27) corresponds to the
famous Einstein's equation
$dm / d\tau$ $=$ $\left(E'_{in}~-~E'_{out}\right) / c^{2}$,
if also Eqs. (3) and (18) are used.

Eqs. (25)-(27) yield for the four-acceleration of the IDP
\begin{eqnarray}\label{28}
\frac{dU^{\mu}}{d\tau} &=& \sigma'_{pr}~\frac{S}{m c}~\frac{\alpha~\omega}{u}~
\left ( \xi^{\mu}~-~\frac{U^{\mu}}{c} \right )~
+~ \sigma'_{tot} ~\frac{S}{m c} ~ \frac{\alpha ~ \omega}{u} ~\sum_{j=2}^{3}
       \frac{\sigma'_{pr, j}}{\sigma'_{tot}}~ \left ( \xi_{j}^{\mu} ~-~
       \frac{U^{\mu}}{c} \right )~.
\end{eqnarray}

In the further treatment we will consider the case
$\sigma'_{pr, j}$ $\equiv$ 0 for $j$ $=$ 1, 2.
As a consequence, the equation of motion will be of the form
\begin{eqnarray}\label{29}
\frac{dp^{\mu}}{d\tau} &=& \frac{1}{c}~ \sigma'_{tot} ~S~\frac{\alpha~\omega}{u}~
\nonumber \\
& & \times \left \{ \frac{\sigma'_{pr}}{\sigma'_{tot}} ~ \xi^{\mu} ~-~
\left [ x' ~-~  \left ( 1 ~-~ \frac{\sigma'_{pr}}{\sigma'_{tot}} \right ) \right ]
~\frac{U^{\mu}}{c} \right \}
\end{eqnarray}
and the four-acceleration will be
\begin{equation}\label{30}
\frac{dU^{\mu}}{d\tau}~=~\frac{\sigma'_{pr} ~S}{m c}~\frac{\alpha~\omega}{u}~
\left(\xi^{\mu}~-~\frac{U^{\mu}}{c}\right)~.
\end{equation}
The change of particle's mass is given by Eq. (27).
Conventional approach is
that the force due to the solar wind bombardment considers fixed mass of the
particle (see, e. g., Mukai and Yamamoto 1982).

\subsection{Total scattering cross section}
We have used the total scattering and pressure cross sections, in the
previous theoretical parts. We need to determine them. We will consider
spherical particle.

We will use a some sort of approximation to the hard-core scattering problem,
which corresponds to the limiting case of a short-range potential
$V(r)$ $=$ $\infty$ for $r < R$,
$V(r)$ $=$ 0 for $r > R$ (hard-core potential, see, e. g., Iro 2002, p. 158).
We will consider the scattering of point solar wind corpuscles
from an almost hard sphere of radius $R$. In the case of the infinitely hard
sphere of radius $R$, "the dynamics reduces to the laws of reflection
at the surface of the sphere" (Iro 2002, p. 158). The result of classical
physics is following:
"In the case of a finite-range potential, the total [scattering] cross
section is finite and gives the effective area of the potential. (This is
actually the definition of a finite-range potential.) For example, when
point masses are incident onto a hard sphere, $\sigma'_{tot}$ is the cross
section of the sphere -- only particles incident within that area are deflected."
(Iro 2002, p. 161).

However, correct physics for the incident electromagnetic radiation
suggests that geometric cross section may not lead to correct results
for the incident solar wind corpuscles (Kla\v{c}ka 2008a, 2008b).
Inspiring by de Broglie's idea about wave character of massive particles,
one can come to the conclusion that scattering by a hard sphere at very
high energies leads to the total scattering cross section
\begin{equation}\label{31}
\sigma'_{tot} = 2 \pi R^{2}
\end{equation}
and "the classical total cross section is just half of the quantum-mechanical
result in the limit of very short wavelength" (Messiah 1999, pp. 393-395).

If we use a some sort of approximation to the hard sphere, we can use
the total scattering cross section given by Eq. (31), in our paper.

As for the comparison of the results obtained by quantum and non-quantum
physics, we will use Eq. (25) or Eq. (29). The non-quantum approach uses
$\sigma'_{tot}$ $=$ $\sigma'_{pr}$ $=$ $A'$ $=$ $\pi R^{2}$:
\begin{eqnarray}\label{32}
\left ( \frac{dp^{\mu}}{d\tau} \right )_{non-quantum} &=&
		  \frac{1}{c}~ A' ~S ~\frac{\alpha~\omega}{u}~
		  \left ( \xi^{\mu} ~-~ x' ~ \frac{U^{\mu}}{c} \right ) ~.
\end{eqnarray}
Comparison between Eqs. (29) and (32) yields (compare coefficients
at $\xi^{\mu}$ and $U^{\mu}$):
\begin{eqnarray}\label{33}
\sigma'_{pr} &=& A' ~,
\nonumber \\
x' \left ( quantum \right ) &=&
		\left [ x' \left ( non-quantum \right ) ~-~ 1 \right ] ~
		\frac{\sigma'_{pr}}{\sigma'_{tot}} ~+~ 1 ~.
\end{eqnarray}
Using also Eq. (31), one obtains
\begin{equation}\label{34}
x' \left ( quantum \right ) = \frac{1}{2} \left [
			      x' \left ( non-quantum \right ) ~+~ 1 \right ] ~.
\end{equation}

The case $\left \{ \sigma'_{pr} = A', ~\sigma'_{tot} = 2 A' \right \}$
is analogous to the cases of perfectly absorbing or reflecting spheres within
geometrical optics approximation for electromagnetic radiation (Kla\v{c}ka 2008b).
This analogy explains also the importance of quantum physics in our derivations
-- non-quantum physics would not yield correct results in the limit
$u$ $\rightarrow$ $c$.

\subsection{Equation of motion to the second order in $v/u$}
In the approximation to the first order in $v/c$, we can replace the
spacelike part of the four-acceleration of the IDP by acceleration
$d\vec{v}/dt$, where $t$ is time measured in the stationary reference
frame (associated with the Sun).
Further, using Eqs. (2), (11), (14), (15), (17) and (22),
we can express the right-hand side of Eq. (30) in the approximation
to the second order in $v/u$. We get
\begin{eqnarray}\label{35}
\frac{d\vec{v}}{dt} &=& \frac{A' n m_{1} u^{2}}{m}~\left\{\left[
\left(1~-~\frac{\vec{v}\cdot\hat{\vec{u}}}{u}\right)~\hat{\vec{u}}
~-~\frac{\vec{v}}{u}\right]~+~\frac{1}{2}~\frac{v^{2}}{u^{2}}~
\hat{\vec{u}}~+~\frac{\vec{v}\cdot\hat{\vec{u}}}{u}~
\frac{\vec{v}}{u}\right\} ~,
\nonumber \\
A' &=& \pi R^{2} = \sigma'_{pr} ~,
\end{eqnarray}
where $\hat{\vec{u}}$ $\equiv$ $\vec{u}/u$ is the unit vector in
direction of the solar wind.

Let us introduce the cylindrical coordinate system associated with
the orbital plane of the IDP and determined by unit vectors
$\vec{e}_{R}$ (radial vector), $\vec{e}_{T}$ (transversal vector) and
$\vec{e}_{N}$ $=$ $\vec{e}_{R} \times \vec{e}_{T}$ (normal vector).
We can write (Kla\v{c}ka 1994)
\begin{equation}\label{36}
\hat{\vec{u}}~=~\gamma_{R} \vec{e}_{R}~+~\gamma_{T} \hat{\vec{u}}_{T}~,
\end{equation}
where
\begin{equation}\label{37}
\gamma_{R}~=~\cos{\varepsilon}~,~~\gamma_{T}~=~\sin{\varepsilon}
\end{equation}
and
\begin{eqnarray}\label{38}
\hat{\vec{u}}_{T} &=& \frac{1}{N}~\vec{k}\times\vec{e}_{R}~=~\frac{1}{N}~\left(\vec{e}_{T} 
\cos{i}~-~\vec{e}_{N} \cos{\Theta} \sin{i}\right)~,
\nonumber\\
N &=& \sqrt{\left(\cos{i}\right)^{2}~+~\left(\cos{\Theta}\right)^{2}
\left(\sin{i}\right)^{2}}~.
\end{eqnarray}
The quantity $\varepsilon$ is an angle between the radial direction and
the real direction of the solar wind. The unit vector $\vec{k}$ 
corresponds to the vector of angular velocity of solar rotation. 
The inclination of the orbital plane of the IDP with respect to the solar equatorial plane is $i$.
Finally, $\Theta$ is a position angle of the IDP (an angle measured
from the ascending node of the orbit of the IDP to its actual position).

Inserting Eqs. (36) and (38) to Eq. (35), and using the decomposition
of the velocity vector into its radial and transversal components,
$\vec{v}$ $=$ $v_{R} \vec{e}_{R} + v_{T} \vec{e}_{T}$, one obtains
\begin{eqnarray}\label{39}
\frac{d\vec{v}}{dt} &=& \frac{A' n m_{1} u^{2}}{m}~ \left \{
X_{R} \vec{e}_{R} ~+~ X_{T} \vec{e}_{T} ~-~ X_{N} \vec{e}_{N} \right .
\nonumber \\
& & \left . +~ \gamma_{R}~\frac{v_{R}}{u}~ \frac{\vec{v}}{u} ~+~
\gamma_{T}~\frac{\cos{i}}{N}~\frac{v_{T}}{u}~ \frac{\vec{v}}{u}~\right \} ~,
\nonumber\\
X_{R} &=& \gamma_{R}~-~\left(1~+~\gamma_{R}^{2}\right) \frac{v_{R}}{u}~-~
\gamma_{R} \gamma_{T}~\frac{\cos{i}}{N}~\frac{v_{T}}{u} ~+~
\gamma_{R}~\frac{1}{2}~\frac{v^{2}}{u^{2}}
\nonumber\\
X_{T} &=& \left ( 1~-~\gamma_{R}~\frac{v_{R}}{u}~-~\gamma_{T}~
\frac{\cos{i}}{N}~\frac{v_{T}}{u}~+~\frac{1}{2}~\frac{v^{2}}{u^{2}}~
\right ) \gamma_{T}~\frac{\cos{i}}{N} ~-~ \frac{v_{T}}{u}
\nonumber\\
X_{N} &=& \left ( 1 ~-~ \gamma_{R}~\frac{v_{R}}{u}~-~
\gamma_{T}~\frac{\cos{i}}{N} ~ \frac{v_{T}}{u}~
+~\frac{1}{2}~\frac{v^{2}}{u^{2}} ~ \right ) \gamma_{T}~
\frac{\cos{\Theta}\sin{i}}{N} ~.
\end{eqnarray}
The angle $\varepsilon$ is small, its value lies between
2$^{\circ}$ - 3$^{\circ}$ (Bruno {\it et al.} 2003). Thus, we
can neglect terms proportional to $\gamma_{T}^{2}$ and $\gamma_{T}
(v/u)^{2}$. Similarly, we put $\gamma_{R}$ $\approx$ $1$. Then
\begin{eqnarray}\label{40}
\frac{d\vec{v}}{dt} &=& \frac{A' n m_{1} u^{2}}{m}~\left\{\left(~
1~-~2~\frac{v_{R}}{u}~-~\gamma_{T}~\frac{\cos{i}}{N}~\frac{v_{T}}{u}~
+~\frac{1}{2}~\frac{v^{2}}{u^{2}}~\right)\vec{e}_{R}
\right.
\nonumber\\
&& \left. +~\left[\left(~1~-~\frac{v_{R}}{u}~\right)
\gamma_{T}~\frac{\cos{i}}{N}~-~\frac{v_{T}}{u}~\right]
\vec{e}_{T} \right .
\nonumber \\
& & \left . ~-~ \left( 1 ~-~ \frac{v_{R}}{u} \right)
\gamma_{T}~\frac{\cos{\Theta}\sin{i}}{N}~\vec{e}_{N}~
+~\frac{v_{R}}{u}~\frac{\vec{v}}{u}~\right\} ~.
\end{eqnarray}

\section{Equation of motion - electromagnetic radiation effect}
Up to now, we dealt with the action of solar wind on the motion of an
IDP. The role of solar electromagnetic radiation cannot be neglected
in the motion of the IDP in the Solar System.
Relativistically covariant equation of motion for an arbitrarily shaped
dust particle under the action of parallel beam of photons 
(Kla\v{c}ka 2008a, 2008b):
\begin{eqnarray}\label{41}
\frac{d p^{\mu}}{d \tau} &=& \sum_{j=1}^{3} \left (
      \frac{w_{1}^{2} ~S_{elmg}~ \bar{C}'_{pr, j}}{c^{2}} ~+~
      \frac{1}{c} ~F'_{e, j} \right )
      \left ( c ~ b_{j}^{\mu} ~-~ U^{\mu}  \right ) ~,
\end{eqnarray}
where $p^{\mu}$ is four-vector of the particle of mass $m$,
four-vector of the world-velocity of the particle is given by Eq. (21)
and four-vectors $b_{j}^{\mu}$, $j$ $=$ 1, 2, 3 are given as:
\begin{eqnarray}\label{42}
b^{\mu}_{j} &=& \left(\frac{1}{w_{j}}~;~\frac{\vec{e}_{j}}{w_{j}}\right)~,
\nonumber\\
w_{j} &=& \gamma \left(v\right) \left(1-\frac{\vec{v}\cdot\vec{e}_{j}}
{c}\right)~,
\nonumber\\
\vec{e}_{j} &=& \left[\gamma \left(v\right) \left(1+
\frac{\vec{v}\cdot\vec{e'}_{j}}{c}\right)\right]^{-1} \left\{
\vec{e'}_{j}+\left[\left(\gamma \left(v\right)-1\right) \frac{\vec{v}\cdot\vec{e'}_{j}}{v^{2}} 
+\frac{\gamma\left(v\right)}{c}\right]\vec{v}\right\}~,
\nonumber\\
&& j~=~1,~2,~3~,
\end{eqnarray}
where $\{\vec{e'}_{j};~j=1, 2, 3\}$ is orthonormal vector basis in the proper 
reference frame of the particle and $\{\vec{e}_{j};~j=1, 2, 3\}$ is 
corresponding vector basis in the stationary frame; $\vec{e}_{1}$ corresponds 
to the radial direction (i.e. the Sun - particle direction). 
$S_{elmg}$ is flux density of the electromagnetic radiation and
$\bar{C}'_{pr, j}$ ($j$ $=$ 1, 2, 3) are spectrally averaged cross sections
of radiation pressure
\begin{eqnarray}\label{43}
\bar{C}'_{pr, j} &=& \frac{\int_{0}^{\infty} I \left ( \lambda \right ) ~
		     C'_{pr, j} \left ( \lambda \right ) ~ d \lambda}{
		     \int_{0}^{\infty} I \left ( \lambda \right ) ~ d \lambda} ~,
		     ~~~j = 1, 2, 3 ~,
\end{eqnarray}
where $I ( \lambda )$ is the flux of monochromatic radiation energy. If
$\bar{C}'_{pr, 2}$ $=$ $\bar{C}'_{pr, 3}$ $\equiv$ 0, then Eq. (41) reduces
to the Poynting-Robertson effect (Poynting 1903, Robertson 1937,
Kla\v{c}ka 2008a, 2008b, Kla\v{c}ka {\it et al.} 2009), since also components 
of the thermal emission
force $F'_{e, j}$ ($j$ $=$ 1, 2, 3) are equal zero, in that case 
(Mishchenko 2001, Mishchenko {\it et al.} 2002). 
It can be easily verified that
Eq. (41) yields $d m / d \tau =$ 0, i.e., mass of the particle
is conserved, under the action of electromagnetic radiation.

To the first order in $\vec{v} / c$, Eq. (41) yields
\begin{eqnarray}\label{44}
\frac{d \vec{v}}{d t} &=&  \frac{S_{elmg}}{m~c} ~ \sum_{j=1}^{3}
      \bar{C}'_{pr, j} ~\left [ \left ( 1~-~ 2~
      \vec{v} \cdot \vec{e}_{1} / c ~+~
      \vec{v} \cdot \vec{e}_{j} / c \right ) ~ \vec{e}_{j}
      ~-~ \vec{v} / c \right ]
\nonumber   \\
& &   +~\frac{1}{m} ~ \sum_{j=1}^{3} F'_{e, j} ~\left [  \left ( 1~+~
      \frac{\vec{v} \cdot \vec{e}_{j}}{c} \right )  \vec{e}_{j}
      ~-~ \frac{\vec{v}}{c} \right ]  ~,
\nonumber \\
\vec{e}_{j} &=& ( 1 ~-~ \vec{v} \cdot \vec{e'}_{j} / c ) ~
\vec{e'}_{j}~+~\vec{v} / c ~~~, ~~j = 1, 2, 3~.
\end{eqnarray}
It is worth mentioning to stress that the values of radiation pressure
cross sections $\bar{C}'_{pr, j}$, $j$ $=$ 1, 2, 3,
depend on particle's orientation with respect to
the incident radiation -- their values are time dependent, in general.
General equation of motion, represented by Eq. (41) or Eq. (44) differs
from the Poynting-Robertson effect. Eqs. (41)-(44)
hold for arbitrarily shaped particles. Experimental evidence that
nonspherical dust grains move in a different way than spherical particles
was given by Krauss and Wurm (2004).

As it is conventionally used in Solar System studies, we will restrict
ourselves to the Poynting-Robertson effect, as for the effect of solar
electromagnetic radiation. Thus, instead of Eq. (44), we will use
\begin{equation}\label{45}
\frac{d \vec{v}}{d t} =  \frac{S_{elmg}~A'~ \bar{Q}'_{pr, 1}}{m~c} ~
      \left \{ \left ( 1~-~
      \frac{\vec{v} \cdot \vec{e}_{1}}{c} \right )  \vec{e}_{1}
      ~-~ \frac{\vec{v}}{c} \right \} ~,
\end{equation}
where a dimensionless efficiency factor of radiation pressure
$\bar{Q}'_{pr, 1}$ is defined by relation
$ \bar{Q}'_{pr, 1}$ $=$ $\bar{C}'_{pr, 1}/A'$. The values of
$\bar{Q}'_{pr, 1}$ can be calculated according to Mie (Mie 1908; see also
van de Hulst 1981, Bohren and Huffman 1983).

\subsection{Electromagnetic radiation effect as a special case of the 
solar wind effect}
The transformations \\
i) $\vec{u}$ $\to$ $c~ \vec{e}_{1}$, \\
ii) $S'$ $=$ $S~ \alpha~ \omega / u$ $\to$
$S'_{elmg}$ $=$ $S_{elmg}~ w_{1}^{2}$ (Kla\v{c}ka 2008b), \\
iii) $\sigma'_{pr,j}$ $\to$ $\bar{C}'_{pr,j}$ ($j$ $=$ 1, 2, 3), and, \\
iv) $x'$ $=$ 1 (Kla\v{c}ka 2008b) \\
reduce Eq. (25) into Eq. (41) 
without thermal emission terms
($F'_{e, j}$ $\equiv$ 0 for $j$ $=$ 1, 2, 3). 
This means that our theory for
electromagnetic radiation effect without thermal emission is consistent
with the theory for corpuscular radiation effect.

\section{Equation of motion -- solar radiation and solar gravity}
Let us consider spherical body orbiting the Sun under the action of solar
radiation, i.e. solar corpuscular (solar wind) and electromagnetic radiation.
The effect of the solar electromagnetic radiation on the motion of
spherical particle corresponds to the Poynting-Robertson effect
(P-R effect).

\subsection{Solar electromagnetic radiation effect}
It is useful to introduce a $\beta$-parameter defined as the ratio of
radial component of the radiation force and the gravitational
force between the Sun and the particle with zero velocity:
\begin{eqnarray}\label{46}
\beta &=& \frac{L_{\odot} ~A' ~\bar{Q}'_{pr}}{4 ~\pi ~c~ m ~\mu} ~,
\nonumber\\
\mu &\equiv& G \left(M_{\odot}+m \right)~\doteq~GM_{\odot}~. 
\end{eqnarray}
$L_{\odot}$ $=$ 3.842 $\times$ 10$^{26}$ W (Bahcall 2002) is luminosity
of the Sun, $\bar{Q}'_{pr}$ $\equiv$ $\bar{Q}'_{pr,1}$, $G$ is the 
gravitational constant and $M_{\odot}$ is mass
of the Sun. For homogeneous spherical particle we can write
\begin{equation}\label{47}
\beta = 5.760 \times 10^{2}
\frac{\bar{Q}'_{pr}}{R [\mu \mbox{m}]~\rho [\mbox{kg/m}^{3}]}~,
\end{equation}
where $R$ is radius of the particle and $\rho$ is mass density of the
particle. Conventionally it is assumed that $\beta$ $=$ $const$:
neither optical properties nor mass of the IDP change. We do not
restrict ourselves to the validity of this assumption.

Now, using Eq. (46), the relation $S_{elmg}$ $=$ $L_{\odot} / (4 \pi r^{2})$
holds, where $r$ is a heliocentric distance of the IDP. On the basis of the
decomposition of the velocity vector
$\vec{v}$ $=$ $v_{R} \vec{e}_{R}$ $+$ $v_{T} \vec{e}_{T}$,
we can rewrite Eq. (45) into the form
\begin{equation}\label{48}
\left(\frac{d\vec{v}}{dt}\right)_{P-R}~=~\beta~\frac{\mu}{r^{2}}~
\left[\left(1~-~2~\frac{v_{R}}{c}\right)\vec{e}_{R}~-~\frac{v_{T}}{c}~
\vec{e}_{T}\right]~,
\end{equation}
This is the acceleration of the IDP under the action of the 
Poynting-Robertson effect.

\subsection{Solar wind effect}
Let us replace the fraction behind the curly braces in Eq. (40) by new
quantities
\begin{equation}\label{49}
\frac{A' n m_{1} u^{2}}{m}~ \equiv ~
      \frac{\eta}{\bar{Q}'_{pr}} ~\beta ~ \frac{u}{c} ~ \frac{\mu}{r^{2}} ~.
\end{equation}
$\eta$ is conventionally a constant of the value: 0.22 (Whipple 1967,
Dohnanyi 1978) or 0.30 (Gustafson 1994, Abe 2009). 
The solar wind speed $u$ values
of 350 km/s (Dohnanyi 1978) or 400 km/s (Gustafson 1994)
are used in modelling of orbital evolution under the action of solar wind.

Let us look on the numerical values of $\eta$ and $u$ on the basis of the
solar physics data. In order to calculate the value of $\eta$, we need the
values of $n$, $u$ and $E_{1}$, on the basis of Eq. (15). The average values
near the orbit of the Earth (1 AU) are (Hundhausen 1997, p. 92):
proton density $n_{1}$ $=$ 6.6 cm$^{-3}$,
electron density $n_{2}$ $=$ 7.1 cm$^{-3}$,
He$^{2+}$ density $n_{3}$ $=$ 0.25 cm$^{-3}$,
flow speed $u$ $=$ 450 km s$^{-1}$. Eq. (15) yields for the average value
$S$ (solar wind; 1 AU) $=$ $u$ $\sum_{i=1}^{3} n_{i} E_{1~i}$ $\doteq$
$c^{2}$ $u$ $\sum_{i=1}^{3} n_{i} m_{i}$ $=$ 515.642 kg s$^{-3}$.
Moreover, we will take into account that $n_{i} u$ (1 AU) $=$
$\langle n_{i} \rangle$ $\langle u \rangle$ (1 AU)
($1 - 0.15 \cos \varphi$)$^{2}$,
$\varphi$ $=$ 2 $\pi$ $[ t - t (max) ]	/ T$, $T$ $=$ 11.1 years and
$t (max)$ is the time of the solar cycle maximum
(Svalgaard 1977 -- chapter 13). This result, together with Eqs. (11)
( $\omega$ $\doteq$ 1), (14), (17) and (30), yields
$S (solar~wind)$ $=$ $S_{elmg}$ ($A' / \sigma_{pr} '$) $\eta$
and
\begin{eqnarray}\label{50}
\eta &=& \eta_{0} ~ \left ( 1 ~-~ \delta~ \cos \varphi \right )^{2} ~,
\nonumber \\
u &=& u_{0} ~ \left ( 1 ~-~ \delta~ \cos \varphi \right ) ~,
\nonumber \\
\eta_{0} &=& 0.38 ~,
\nonumber \\
\delta &=& 0.15 ~,
\nonumber \\
u_{0} &=& 450~ \mbox{km/s} ~,
\nonumber \\
\varphi &=& 2 \pi ~\frac{t - t_{retard} - t \left ( max \right )}{T}  ~,
\nonumber \\
T &=& 11.1 ~ \mbox{years} ~,
\end{eqnarray}
if we put $\sigma_{pr}$ $=$ $A'$ and $S_{elmg} (1 AU)$ $=$
$L_{\odot}$ $/$ [ 4 $\pi$ $[ r(=1 AU) ]^{2}$ ],
$L_{\odot}$ $=$ 3.842 $\times$ 10$^{26}$ W. The value of $T$
represents the average value of the solar cycle period (see, e. g.,
Foukal 2004, p. 366).
The retarded time $t_{retard}$ is of the order
$r / u_{0}$ and it is only a better approximation to reality
than the omission of this term. 

\subsubsection{More exact solution}
In reality, one needs to know
concentration $n (r, t)$ and solar wind velocity $u (r, t)$, when the
dust grain is situated at a position of heliocentric distance $r$ at
the time $t$. More precise information can be obtained from
continuity equation (radial component of the velocity $\vec{u}$ is 
approximated by the magnitude $u$)
\begin{equation}\label{51}
\frac{\partial n}{\partial t} ~+~ \frac{1}{r^{2}} ~
\frac{\partial \left ( r^{2} n u  \right )}{\partial r} = 0 ~, 
\end{equation}
if $r \ne$ 0. Using the observational fact $n$ $=$ $const$ $u$ / $r^{2}$
(Svalgaard 1977 -- chapter 13), one obtains
\begin{equation}\label{52}
\frac{\partial u}{\partial t} ~+~
\frac{\partial  u^{2}}{\partial r} = 0 ~, 
\end{equation}
if $r \ne$ 0. Using the boundary condition 
\begin{equation}\label{53}
\lim_{r \rightarrow 0} u ( r, t ) = u_{0}  \left \{ 1 ~-~ \delta~ 
\cos \left [  2 \pi ~ \frac{ t -  t (max)}{T} \right ]  \right \} ~, 
\end{equation}
the quasi-linear partial differential equation Eq.(52) can be solved.

Eq. (52) is known as Burgers equation: \\
$\partial u/\partial t$ $+$
$u$ $\partial u/\partial x$ = 0 ~,\\ 
see, e.g., \v{S}ev\v{c}ovi\v{c} (2008, pp. 40 - 42). \\
If the boundary condition \\
$u ( x = 0, t)$ $=$ $\chi (t)$ \\
is given, then the solution of the Burgers equation is \\
$u ( x, t)$ $=$ $\chi ( t - x / u ( x, t) )$. \\
The last nonlinear algebraic equation can be solved 
by an iteration method, e. g.: \\
$u$ $=$ $\chi ( t - x / u_{0} )$ $+$ $\lim_{k \rightarrow \infty}$
$v_{k}$, \\
$v_{k+1}$ $=$  
$\chi \{ t - x / [ \chi ( t - x / u_{0} ) + v_{k} ] \}$ $-$
$\chi ( t - x / u_{0} )$, \\
$v_{1}$ $=$ 0. 

On the basis of the known solution of the Burgers equation,
Eqs. (52)-(53) yield 
\begin{equation}\label{54}
u ( r, t ) = u_{0}  \left \{ 1 ~-~ \delta~ \cos
\left [  2 \pi ~ \frac{t - r / \left ( 2 u \left ( r, t \right ) 
\right ) - t (max)}{T} \right ]  \right \} ~. 
\end{equation} 
Comparison with Eq. (50) shows that the retarded time is $t_{retard}$ 
$=$ $r / ( 2 u )$. 

The nonlinear algebraic Eq. (54) can be solved by the iteration method 
presented above, or by the following iteration: 
\begin{equation}\label{55}
u_{k+1} ( r, t ) = u_{0}  \left \{ 1 ~-~ \delta~ \cos
\left [  2 \pi ~ \frac{t - r / \left ( 2 u_{k}  \left ( r, t \right )
\right ) - t (max)}{T} \right ]  \right \} ~,
\end{equation} 
since the right-hand side of Eq. (54) is a contractive/contraction 
function for the case $u^{2}$ $>$ $\pi$ $u_{0}$ $r$ $\delta$ $/$ $T$ 
and heliosphere is characterized by condition $r$ $<$ 150 AU 
(approximately). We can put $u_{1} ( r, t )$ $=$ $u_{0}$.

\subsubsection{Summary}
Using defition by Eq. (49), we can summarize, on the basis of Eqs. (50)
and (54):
\begin{eqnarray}\label{56}
\frac{A' n m_{1} u^{2}}{m}~ &\equiv& ~
      \frac{\eta}{\bar{Q}'_{pr}} ~\beta ~ \frac{u}{c} ~ \frac{\mu}{r^{2}} ~, 
\nonumber \\
u ( r, t ) &=& u_{0}  \left \{ 1 ~-~ \delta~ \cos
\left [  2 \pi ~ \frac{t - r / \left ( 2 u \left ( r, t \right ) 
\right ) - t (max)}{T} \right ]  \right \} ~, 
\nonumber \\
\eta (r, t) &=& \eta_{0} [u(r, t) / u_{0}]^{2} ~,
\nonumber \\
\eta_{0} &=& 0.38 ~,
\nonumber \\
\delta &=& 0.15 ~,
\nonumber \\
u_{0} &=& 450~ \mbox{km/s} ~,
\nonumber \\
T &=& 11.1~ \mbox{years} ~.
\end{eqnarray} 
One can use also $n$ $\equiv$ $n(r, t)$ $=$
$n_{0}$ [$u(r, t) / u_{0}$] (1 AU $/$ $r$ [AU])$^{2}$.
These results represent a more realistic model than the model
conventionally used. It takes into account more observational
facts (Svalgaard 1977, Hundhausen 1997).

A more simple accesses use Eq. (50) with 
$t_{retard}$ $=$ $r$ $/$ (2$u_{0}$) or $t_{retard}$ $=$ 0.
Shock waves do not exist in these simple cases. The shock waves 
are generated by solution of the Burgers equation and this is
considered in Eq. (56). As was pointed out in the comment to Eq. (55), 
the shock waves are not realized in the Solar System.

We put $i$ $=$ 0 in Eq. (40).
Then, the acceleration of the IDP caused by the solar wind has the form
\begin{eqnarray}\label{57}
\left( \frac{d\vec{v}}{dt}\right)_{SW} &=& \frac{\eta}{\bar{Q}'_{pr}} ~\beta~
	  \frac{\mu}{r^{2}}~ \biggl \{
	  \left ( \frac{u}{c}~-~2~\frac{v_{R}}{c}
	  \right ) \vec{e}_{R}~-~\frac{v_{T}}{c}~\vec{e}_{T}
\nonumber\\
& &
-~\gamma_{T}~\left [ \frac{v_{T}}{c}~\vec{e}_{R}~-~
\left(\frac{u}{c} ~-~ \frac{v_{R}}{c} \right) \vec{e}_{T} \right ]
\nonumber\\
& & + ~\frac{1}{2}~\frac{v^{2}}{uc}~\vec{e}_{R}~+~\frac{v_{R}}{c}~
\frac{\vec{v}}{u} \biggr \}~.
\end{eqnarray}

\subsection{Equation of motion}
Gravitational acceleration from the Sun is $-(\mu/r^{2})\vec{e}_{R}$.
Neglecting the solar wind pressure term $\beta (\eta/\bar{Q}'_{pr})
(\mu/r^{2}) (u/c)$ in Eq. (57), we can write the final equation of
motion of the IDP in the form
\begin{eqnarray}\label{58}
\frac{d\vec{v}}{dt} &=& -~\frac{\mu~\left ( 1 - \beta \right )}{r^{2}}~
\vec{e}_{R}~-~ \left( 1~+~\frac{\eta}{\bar{Q}'_{pr}} \right) \beta ~
\frac{\mu}{r^{2}}~ \left(2~\frac{v_{R}}{c}~\vec{e}_{R}~+~\frac{v_{T}}{c}~
\vec{e}_{T} \right )
\nonumber \\
& & + ~ \frac{\eta}{\bar{Q}'_{pr}} ~\beta ~\frac{\mu}{r^{2}} \biggl \{
-~\gamma_{T}~\frac{v_{T}}{c}~\vec{e}_{R}~+~\gamma_{T}~\left(
\frac{u}{c}~-~\frac{v_{R}}{c}\right)\vec{e}_{T}
\nonumber\\
&& +~\frac{1}{2}~\frac{v^{2}}{uc}~\vec{e}_{R}~+~
\frac{v_{R}}{c}~\frac{\vec{v}}{u} \biggr \}~;
\end{eqnarray}
Eqs. (48) and (57) were used.
We introduced a new central acceleration
$-~\mu (1-\beta) \vec{e}_{R} /r^{2}$, i.e.
the gravitational acceleration from the Sun reduced by the solar
electromagnetic radiation pressure. Other terms on the right-hand side
of Eq. (58) constitute the nongravitational disturbing acceleration.
Eq. (50) has to be taken into account.

Moreover, Eq. (27) can be rewritten to the form describing decrease of
particle's radius $R$:
\begin{eqnarray}\label{59}
\frac{dR}{dt} &=& -~ \frac{K}{r^{2}} ~\frac{|\vec{u} - \vec{v}|}{u} ~
\left ( 1 - \delta \cos \varphi \right ) ^{2}  ~,
\nonumber \\
u &=& u_{0} ~ \left ( 1 ~-~ \delta~ \cos \varphi \right ) ~,
\nonumber \\
\delta &=& 0.15 ~,
\nonumber \\
u_{0} &=& 450~ \mbox{km/s} ~,
\nonumber \\
\varphi &=& 2 \pi ~\frac{t - t_{retard} - t \left ( max \right )}{T}  ~,
\nonumber \\
T &=& 11.1 ~ \mbox{years} ~,
\end{eqnarray}
where $K$ is a constant characterizing decrease of the radius of the particle;
Eq. (50) was used, also (see also text between Eqs. 50 and 57).
Eq. (37) has to be used, too: $\vec{u}$ $=$ $u$
$\hat{\vec{u}}$. One can use Eq. (59) as an approximation to the process of
erosion of the particle due to the solar wind corpuscles.
Since $\beta$ is a function of $R$ (also $\bar{Q}'_{pr}$ is a function of $R$),
Eqs. (50), (58)-(59) have to be solved simultaneously, together with the
Mie's calculations yielding $\bar{Q}'_{pr}$ for a given $R$.

The real flux density of solar wind energy and the approximation of a 
constant flux ($\eta$ $=$ $\eta_{0}$) for the radial solar wind, 
can be described by the approximation
\begin{equation}\label{60}
\eta \doteq \eta_{0} = 0.38 ~.
\end{equation}

Analytical approach to solution of Eqs. (50), (58)-(59) is presented in Sec. 5.
Detail numerical solutions are given in Sec. 6.

\section{Secular evolution of particle's orbital elements
under the action of solar radiation -- analytical approach}
We have obtained complete equation of motion in the preceding section.
We want to obtain qualitative understanding of the orbital evolution
of IDP. This task is the main subject of this section.
Trend is that $\beta$ is a decreasing function of $R$.
Thus, the term $-$ [$\mu (1 - \beta) / r^{2}$] $\vec{e}_{R}$, in Eq. (58),
does not correspond to Keplerian acceleration, if Eq. (59) is
taken into account. As a consequence, osculating orbital elements have to
be calculated for Keplerian acceleration and it is given by the term
$-$ ($\mu / r^{2}$) $\vec{e}_{R}$.

We will calculate the secular evolution of semimajor axis $a$, eccentricity $e$
of the particle's orbit under this nongravitational perturbation.
We have to use Eqs. (100) or (103) in Kla\v{c}ka (2004) and Eqs. (32), (34)
and (37) in Kla\v{c}ka (1993b), for the
time evolution of the osculating elements.

Secular values of semimajor axis and eccentricity, when central
acceleration is given by solar gravity term $-$ ($\mu / r^{2}$) 
$\vec{e}_{R}$, are
\begin{eqnarray}\label{61}
a &=& a_{\beta} ~( 1 - e_{\beta}^{2} )^{3/2} ~ \frac{1}{2 \pi} ~
	       \int_{0}^{2 \pi} \frac{ \left [ 1 +\beta  \left ( 1 +
	       e_{\beta}^{2} + 2 e_{\beta} \cos{x} \right ) /
	       \left ( 1 - e_{\beta}^{2} \right ) \right ]^{-1}}{
	       \left ( 1~+~e_{\beta} \cos{x} \right )^{2}} ~ dx
\nonumber \\
e &=& ( 1 - e_{\beta}^{2} )^{3/2} ~ \frac{1}{2 \pi} ~\int_{0}^{2 \pi}
      \frac{\sqrt{\left ( 1 - \beta \right)^{2} e_{\beta}^{2} +
      \beta^{2} - 2 \beta \left ( 1 - \beta \right) e_{\beta}
      \cos{x}}}{\left( 1~+~e_{\beta} \cos{x} \right)^{2}} ~ dx	~.
\end{eqnarray}
The quantities denoted by index $\beta$ hold for central acceleration
$-$ [$\mu ( 1 - \beta) / r^{2}$] $\vec{e}_{R}$ and it is supposed that
their values do not significantly change during particle's revolution
around the Sun. Analytical approach for secular evolution of orbital
elements is possible under such an assumption, otherwise detail numerical
integration of Eq. (58) has to be done. The quantities denoted by index
$\beta$ can be calculated from Eq. (67) below.

\subsection{Calculation of quantities for Eq. (61)}
Let us use perturbation equations of celestial mechanics in the following form:
\begin{eqnarray}\label{62}
\frac{da_{\beta}}{dt} &=& \frac{a_{\beta}}{1-e_{\beta}^{2}} ~
	       \left \{ 2 ~
	       \sqrt{\frac{p_{\beta}}{\mu \left ( 1 - \beta \right)}}
	       ~\left [ a_{R}~e_{\beta} \sin{f_{\beta}} ~+~
	       a_{T} \left( 1~+~e_{\beta} \cos{f_{\beta}} \right) \right]
	       \right .
\nonumber \\
& & \left .  +~\frac{\dot{\beta}}{1 - \beta} ~\left (
	       1 + e_{\beta}^{2} ~+~ 2~ e_{\beta} \cos{f_{\beta}} \right )
	       \right \}~,
\nonumber \\
\frac{de_{\beta}}{dt}&=& \sqrt{\frac{p_{\beta}}{\mu\left(1-\beta\right)}}
			 ~\left[ a_{R}\sin{f_{\beta}}~+~
	       a_{T} \left(\cos{f_{\beta}}~+~\frac{e_{\beta}+\cos{f_{\beta}}}
	      {1 + e_{\beta} \cos{f_{\beta}}} \right) \right]
\nonumber \\
& &	       +~\frac{\dot{\beta}}{1 - \beta} ~\left (
	       e_{\beta} ~+~ \cos{f_{\beta}} \right ) ~,
\nonumber \\
\frac{d\omega_{\beta}}{dt} &=& -~\sqrt{\frac{p_{\beta}}{\mu\left(1-\beta\right)}}~
    \frac{1}{e_{\beta}}~\left[a_{R}\cos{f_{\beta}}~-~a_{T}\sin{f_{\beta}}
    ~\frac{2+e_{\beta}\cos{f_{\beta}}}
{1+e_{\beta}\cos{f_{\beta}}}\right]
\nonumber \\
& &	     +~ \frac{1}{e_{\beta}}~\frac{\dot{\beta}}{1 - \beta} ~
	     \sin{f_{\beta}}  ~,
\end{eqnarray}
where $p_{\beta}$ $=$ $a_{\beta} (1-e_{\beta}^{2})$, $f_{\beta}$ is
true anomaly of the IDP and $\omega_{\beta}$ is argument of perihelion of the
particle's orbit; the dot over $\beta$ denotes differentiation with respect to
time. It is assumed that the longitude of the ascending node is time
independent. Moreover,
$a_{R}$ and $a_{T}$ are radial and transversal components of the
disturbing acceleration. Eqs. (62) are consistent with Kla\v{c}ka 
(1993a -- Eqs. 12, 14) and  Kla\v{c}ka (1993b -- Eqs. 32, 34, 37 if
$M$ $=$ $M_{\odot} ( 1 - \beta )$).
Using Eq. (58) and expressions
\begin{equation}\label{63}
v_{R}~=~\sqrt{\frac{\mu \left(1-\beta\right)}{p_{\beta}}}~e_{\beta}\sin{f_{\beta}}~,~~
v_{T}~=~\sqrt{\frac{\mu \left(1-\beta\right)}{p_{\beta}}}~\left(1+e_{\beta}
\cos{f_{\beta}} \right)~,
\end{equation}
we obtain
\begin{eqnarray}\label{64}
a_{R} &=& -~ 2 \left( 1 + \frac{\eta}{\bar{Q}'_{pr}} \right) \beta ~
\frac{\mu}{r^{2}}~\frac{\sqrt{\mu \left( 1 - \beta \right) / p_{\beta}}}{c} ~
e_{\beta} \sin{f_{\beta}}
\nonumber \\
& & +~ \frac{\eta}{\bar{Q}'_{pr}} ~\beta ~ \frac{\mu}{r^{2}}~ \frac{1}{c}~
\biggl \{-~\gamma_{T}~\sqrt{\frac{\mu \left(1-\beta\right)}{p_{\beta}}}~\left(
 1 + e_{\beta} \cos{f_{\beta}} \right)
\nonumber\\
&& +~ \frac{\mu \left ( 1 - \beta \right ) / p_{\beta}}{u} ~\left [
~\frac{1}{2}~ \left(1+e_{\beta}^{2}+2e_{\beta} \cos{f_{\beta}}\right)~+~
e_{\beta}^{2}\sin^{2}{f_{\beta}} \right ] \biggr \} ~,
\nonumber \\
a_{T} &=& -~ \left ( 1 + \frac{\eta}{\bar{Q}'_{pr}} \right) \beta ~
\frac{\mu}{r^{2}}~\frac{\sqrt{\mu \left(1-\beta\right) / p_{\beta}}}{c}~
\left ( 1 + e_{\beta} \cos{f_{\beta}} \right )
\nonumber \\
& & +~ \frac{\eta}{\bar{Q}'_{pr}} ~\beta ~
      \frac{\mu}{r^{2}}~\frac{1}{c}~
\biggl \{~\gamma_{T}~u~-~\gamma_{T}~\sqrt{\frac{\mu \left(1-\beta\right)}{
	p_{\beta}}} ~e_{\beta} \sin{f_{\beta}}
\nonumber\\
&& + ~ \frac{\mu \left ( 1 - \beta \right ) / p_{\beta}}{u} ~e_{\beta}
\left ( 1 + e_{\beta} \cos{f_{\beta}} \right) ~\sin{f_{\beta}} \biggr \}~.
\end{eqnarray}
Secular evolution of the orbital element $g$ we get by the time averaging
of $dg/dt$ over one orbital period $P$, i.e.
\begin{equation}\label{65}
\left\langle \frac{dg}{dt} \right\rangle ~ \equiv ~ \frac{1}{P}~
\int_{0}^{P} \frac{dg}{dt}~dt~=~\frac{1}{a_{\beta}^{2}\sqrt{1-e_{\beta}^{2}}}~
\frac{1}{2\pi}~\int_{0}^{2\pi} r^{2}~\frac{dg}{dt}\left(f_{\beta} \right)
~df_{\beta}~.
\end{equation}
We have used the second and the third Kepler's laws:
$r^{2} d f_{\beta} / dt$ $=$ $\sqrt{\mu ( 1 - \beta ) p_{\beta}}$ $-$
$d \omega_{\beta} / dt$ $-$ $( d \Omega_{\beta} / d t )$ $\cos i_{\beta}$
$\doteq$ $\sqrt{\mu ( 1 - \beta ) p_{\beta}}$
and $a_{\beta}^{3} / P^{2}$ $=$ $\mu ( 1 - \beta ) / (4 \pi ^{2} )$;
$p_{\beta}$ $=$ $a_{\beta} (1-e_{\beta}^{2})$, $\omega_{\beta}$ is
argument of perihelion and $\Omega_{\beta}$
is longitude of the ascending node.

As for the quantity $\dot{\beta}$, it is a function of radius $R$ of the
particle. Using an approximation $\beta$ $=$ $A/R$ $+$ $B$, where
$A$ and $B$ are constants for a given particle, we can write
\begin{eqnarray}\label{66}
\beta &=& \frac{A}{R} ~+~ B ~,
\nonumber \\
\frac{d\beta}{dt} &=& ~-~ \frac{A}{R^{2}} ~\frac{dR}{dt}
		   = \frac{A}{R^{2}} ~\frac{K}{r^{2}} ~,
\end{eqnarray}
if also dominant part of Eq. (59) is used ($\delta$ $=$ 0,
$|\vec{u} - \vec{v}|$ $\doteq$ $u$).

Application of	Eq. (65) to Eqs. (62) and (66), using the assumption
that orbital elements and $\beta$-parameter do not significantly change
during particle's revolution around the Sun, yields ($\delta$ $\equiv$ 0
is assumed)
\begin{eqnarray}\label{67}
\frac{da_{\beta}}{dt}  &=& -~\beta~\frac{\mu}{c}~
\frac{2 + 3 e_{\beta}^{2}}{a_{\beta} \left(1-e_{\beta}^{2}\right)^{3/2}}
\nonumber \\
& & \times~ \left \{ 1~+~\frac{\eta_{0}}{\bar{Q}'_{pr}}~ \left [ ~1 ~-~
	2 \gamma_{T} ~\frac{1}{2+3e_{\beta}^{2}}~ \frac{u_{0}}{
	\sqrt{\mu \left ( 1 - \beta \right ) / p_{\beta}}}
	~\right ] \right \}
\nonumber\\
& & +~ \frac{1}{1 - \beta} ~ \frac{A}{R^{2}} ~K ~
   \frac{1 + e_{\beta}^{2}}{a_{\beta} \left ( 1 - e_{\beta}^{2} \right )^{3/2}} ~,
\nonumber \\
\frac{de_{\beta}}{dt} &=& - \beta~\frac{\mu}{c}~
	\frac{5e_{\beta}/2}{a_{\beta}^{2}\sqrt{1-e_{\beta}^{2}}}
\nonumber \\
& & \times~
 \left \{ 1~+~ \frac{\eta_{0}}{\bar{Q}'_{pr}}~\left[~1 ~-~\frac{2}{5} ~
 \gamma_{T} ~\frac{1-\sqrt{1-e_{\beta}^{2}}}{e_{\beta}^{2}}~\frac{u_{0}}{
  \sqrt{\mu \left ( 1 - \beta \right ) / p_{\beta}}} ~\right ] \right \}
\nonumber \\
& & +~ \frac{1}{1 - \beta} ~ \frac{A}{R^{2}} ~K ~
       \frac{e_{\beta}}{a_{\beta}^{2} \sqrt{1-e_{\beta}^{2}}} ~,
\nonumber \\
\frac{d\omega_{\beta}}{dt} &=& - ~\frac{\eta_{0}}{\bar{Q}'_{pr}}~ \beta ~
	    \frac{\mu}{c}~ \frac{1}{a_{\beta}^{2}\sqrt{1-e_{\beta}^{2}}} ~
\nonumber \\
& & \times \left \{
  \gamma_{T}~ \frac{1 - \sqrt{1-e_{\beta}^{2}}}{e_{\beta}^{2}}	~-~
  \frac{1}{2} ~\frac{\sqrt{\mu \left ( 1 - \beta \right ) / p_{\beta}}}{u_{0}}
  \right \} ~,
\end{eqnarray}
where the symbol $\langle ~\rangle$ was omitted on the left-hand sides.
The quantities with index $\beta$ in Eq. (67) are the quantities 
which have to be used in Eq. (61). Eqs. (67) hold under
the assumption $\eta \equiv \eta_{0}$.

\subsection{Complete set of differential equations for
secular evolution of particle's orbital elements}
We have obtained the following set of differential equations for
secular evolution of orbital elements: Eqs. (61), (66)-(67).
This set has to be completed by equation for secular evolution of
radius of the particle. Using Eqs. (59) and (65), we obtain
($\delta$ $=$ 0, $|\vec{u} - \vec{v}|$ $\doteq$ $u$)
\begin{equation}\label{68}
\frac{dR}{dt}~=~-~ \frac{K}{a_{\beta}^{2} \sqrt{1-e_{\beta}^{2}}} ~ .
\end{equation}

Initial conditions must be added to the set of differential
equations for orbital elements. If the particle is ejected from
a parent body of known orbital elements, then the particle's initial
orbital elements have to be calculated from Eqs. (60)-(61) in
Kla\v{c}ka (2004).

If one would take into account also thermal change of optical properties of the
spherical dust particle, then also further changes of orbital elements
exist (secular changes of semimajor axis and eccentricity, perihelion motion).
This would correspond to the change of parameter $A$ and $B$ in Eq. (66).
We do not deal with this case (see Kla\v{c}ka {\it et al.} 2007,
P\'{a}stor {\it et al.} 2009).

\subsection{Discussion}
Let us look on secular evolution of semimajor axis $a_{\beta}$
of the particle's orbit. If the P-R effect and radial velocity component
of the solar wind are considered alone (i.e. $\gamma_{T}$ $\equiv$ 0
and $\dot{\beta}$ $\equiv$ 0), then Eqs. (61) and (66)-(68) show that $a$
is a decreasing function of time. If one takes into account also the
non-radial velocity component of the solar wind, or, $\dot{\beta}$ $>$ 0,
the situation may be different. The secular value of $a$ can be
an increasing function of time. Thus, the effect of real solar wind may cause
particle's spiralling outward from the Sun.
(Similarly, also the secular value of $e$ can be
an increasing function of time.)

\subsection{Radial solar wind and decrease of particle's radius}
According to Eqs. (27) and (59), the mass of the particle may decrease.
Eq. (68) holds under assumption that the decrease of the particle's radius
is small enough during particle's revolution around the Sun.

When we consider only the P-R effect and radial solar wind effect
($\gamma_{T}$ $=$ 0), we are able to calculate the radius of IDP as a
function of its orbital eccentricity. Using Eqs. (67) and (68), one can
immediately write
\begin{eqnarray}\label{69}
\frac{de_{\beta}}{dR} &=& \frac{e_{\beta}}{K} ~ \left \{ \frac{5}{2} ~
	\left ( 1 + \frac{\eta_{0}}{\bar{Q}'_{pr}} \right ) ~ \beta ~
	\frac{\mu}{c} ~-~
	\frac{1}{1 - \beta} ~ \frac{A}{R^{2}} ~K \right \} ~.
\end{eqnarray}
Using also Eqs. (47) and (66), Eq. (69) can be integrated:
\begin{eqnarray}\label{70}
e_{\beta} &=& e_{\beta~in} ~ \left \{
      \frac{A - \left ( 1 - B \right ) R_{in}}{A - \left ( 1 - B \right ) R} ~
      \left ( \frac{R}{R_{in}} \right ) ^{1 + k_{2}} ~ \times
      \exp \left [ k_{1} ~\left ( R - R_{in} \right ) \right ] \right \} ~,
\nonumber \\
k_{1} &=& \frac{5}{2} ~\frac{\mu}{c} ~
\frac{B}{K[\mbox{cm~ AU}^{2}~ \mbox{yr}^{-1}]} ~,
\nonumber \\
k_{2} &=& \frac{5}{2} ~\frac{\mu}{c} ~
\frac{1}{K[\mbox{cm~ AU}^{2}~ \mbox{yr}^{-1}]} \left \{
	  A + \eta \times \frac{5.760 \times 10^{2}}{\varrho \left [
	  \mbox{kg/m}^{3} \right ]} \right \}
\nonumber \\
\beta (R) &=& \frac{A}{R [\mu \mbox{m}]} ~+~ B ~,
\nonumber \\
\mu &=& 4 \pi^{2} \mbox{AU}^{3}~ \mbox{yr}^{-2} ~,
\nonumber \\
c &=& 6.3114 \times 10^{4} \mbox{AU}~ \mbox{yr}^{-1} ~,
\nonumber \\
\eta &\equiv& \eta_{0} = 0.38  ~.
\end{eqnarray}
and the quantities $A$ and $R$ are given in microns; the subscript "in"
denotes initial values.

In order to find secular evolution of semimajor axis $a$ and eccentricity $e$,
according to Eq. (61), we have to solve the following set of equations:
Eq. (66) for a given values of $A$ and $B$,
equation for $a_{\beta}$ in Eq. (67), Eq. (68) and Eq. (70) (or, equation
for $e_{\beta}$ in Eq. 67 instead of Eq. 70).

\subsection{Semi-latus rectum, time of spiralling for radial solar wind and
$K$ $\equiv$ 0}
From Eqs. (67) we can determine also the secular evolution of the
semi-latus rectum $p_{\beta}$ of the particle's orbit. Whence $p_{\beta}$ $=$
$a_{\beta} (1-e_{\beta}^{2})$ we can write
\begin{equation}\label{71}
\left\langle \frac{dp_{\beta}}{dt} \right\rangle ~=~
\left(1-e_{\beta}^{2}\right)\left\langle \frac{da_{\beta}}{dt} \right\rangle~-~
2~a_{\beta} e_{\beta}\left\langle \frac{de_{\beta}}{dt} \right\rangle~,
\end{equation}
what using Eqs. (67)
yields (the symbol $\langle ~ \rangle$ is omitted)
\begin{eqnarray}\label{72}
\frac{dp_{\beta}}{dt} &=&
\frac{\left(1-e_{\beta}^{2}\right)^{3/2}}{p_{\beta}}
\nonumber \\
& & \times ~ \left \{ -~ 2 \left ( 
1 + \frac{\eta_{0}}{\bar{Q'}_{pr}} \right) ~\beta ~\frac{\mu}{c}~
+~ \frac{1}{1 - \beta} ~ \frac{A}{R^{2}} ~K \right \}
\nonumber \\
&& +~2 ~\gamma_{T}~ \frac{\eta_{0}}{\bar{Q'}_{pr}} ~ \beta~
\frac{\mu}{c}~\frac{u_{0}}{\sqrt{\mu \left ( 1 - \beta \right ) / p_{\beta}}}~
\frac{1-e_{\beta}^{2}}{p_{\beta}}~.
\end{eqnarray}
Let us rewrite equation for $\langle de_{\beta}/dt \rangle$ into the form
(the symbol $\langle ~ \rangle$ is omitted)
\begin{eqnarray}\label{73}
\frac{de_{\beta}}{dt} &=&
\frac{e_{\beta} \left ( 1 - e_{\beta}^{2}\right)^{3/2}}{p_{\beta}^{2}}
\nonumber \\
& & \times \left \{ -~\frac{5}{2}~ \left ( 1 +
\frac{\eta_{0}}{\bar{Q'}_{pr}} \right) ~ \beta ~\frac{\mu}{c}~
+~ \frac{1}{1 - \beta} ~ \frac{A}{R^{2}} ~K \right \}
\nonumber \\
& & +~\gamma_{T}~\frac{\eta_{0}}{\bar{Q}'_{pr}} ~\beta~
\frac{\mu}{c}~\frac{u_{0}}{\sqrt{\mu \left(1-\beta\right) / p_{\beta}}}
\nonumber\\
&& \times~
\frac{\left(1-\sqrt{1-e_{\beta}^{2}}\right)\left(1-e_{\beta}^{2}\right)^{3/2}}
{p_{\beta}^{2}~e_{\beta}}~.
\end{eqnarray}

Now, let us consider only the P-R effect and the radial solar wind effect.
We put $\gamma_{T}$ $=$ 0 in Eqs. (72)-(73). In this case, we obtain
the following equation
\begin{eqnarray}\label{74}
\frac{dp_{\beta}}{de_{\beta}} &=& \frac{p_{\beta}}{e_{\beta}}~
   \left \{ -~ 2
   \left ( 1 + \frac{\eta_{0}}{\bar{Q'}_{pr}}\right) \beta ~ \frac{\mu}{c}~
+~ \frac{1}{1 - \beta} ~ \frac{A}{R^{2}} ~K \right \}
\nonumber\\
& & \times \left \{ -~\frac{5}{2}~ \left ( 1 +
    \frac{\eta_{0}}{\bar{Q'}_{pr}} \right) \beta ~\frac{\mu}{c} ~+~
    \frac{1}{1 - \beta} ~ \frac{A}{R^{2}} ~K \right \} ^{-1}
\end{eqnarray}
from Eqs. (72) and (73). Eq. (74) yields the relation
\begin{equation}\label{75}
p_{\beta}~=~p_{\beta~ in} \left(\frac{e_{\beta}}{e_{\beta ~in}}\right)^{4/5}~,
	    ~~~ K \equiv 0 ~,
\end{equation}
where $p_{\beta ~in}$ and $e_{\beta ~in}$ are initial values of semi-latus
rectum and eccentricity of the particle's orbit. Eq. (75) can be considered as a
generalization of the result obtained by Wyatt and Whipple (1950):
we have taken into account not only the P-R effect, but also the radial
solar wind effect. Eq. (75) allows us to write equation for secular
evolution of eccentricity in the form
\begin{eqnarray}\label{76}
\left\langle \frac{de_{\beta}}{dt} \right\rangle &=&
-~\frac{5}{2}~ \left( 1 + \frac{\eta_{0}}{\bar{Q'}_{pr}} \right ) \beta ~
\frac{\mu}{c}~\frac{e_{\beta~ in}^{8/5}}{p_{\beta~ in}^{2}}~
\frac{\left ( 1 - e_{\beta}^{2} \right )^{3/2}}{e_{\beta}^{3/5}} ~,
\nonumber \\
K &\equiv& 0 ~.
\end{eqnarray}
It is evident from Eqs. (74) that due to the P-R effect and radial solar
wind the particle is spiralling inward to the Sun, for $K \equiv$ 0.
Semimajor axis $a_{\beta}$ and eccentricity $e_{\beta}$ of the particle's
orbit converge to 0 (see Eqs. 75, 76 and relation among $a_{\beta}$,
$p_{\beta}$ and $e_{\beta}$).
Eq. (76) can offer the time of spiralling of the particle with initial
orbital elements $a_{\beta~ in}$ and $e_{\beta~ in}$ into the orbit with
osculating elements $a_{\beta}$, $e_{\beta}$. This time is given by relation
\begin{eqnarray}\label{77}
\tau\left(e_{\beta~ in}, e_{\beta} \right) &=&
-~\frac{2}{5}~\left[\left(1+\frac{\eta_{0}}{\bar{Q'}_{pr}} \right ) \beta~
\frac{\mu}{c}\right]^{-1}~a_{\beta~ in}^{2}~\frac{\left(1-e_{\beta~ in}^{2}
\right)^{2}}{e_{\beta~ in}^{8/5}}~I\left(e_{\beta~ in}, e_{\beta} \right)~,
\nonumber\\
I\left(e_{\beta~ in}, e_{\beta} \right) &=& \int_{e_{\beta~ in}}^{e_{\beta}}
\frac{x^{3/5}}{\left(1-x^{2}\right)^{3/2}}~dx~,
\nonumber \\
K &\equiv& 0 ~.
\end{eqnarray}
The time of spiralling of the particle into the Sun is then
\begin{eqnarray}\label{78}
\tau\left(e_{\beta~ in}, 0 \right) &=&
-~\frac{2}{5}~\left[ \left(1+\frac{\eta_{0}}{\bar{Q'}_{pr}} \right ) \beta~
\frac{\mu}{c}\right]^{-1}~a_{\beta~ in}^{2}~\frac{\left(1-e_{\beta~ in}^{2}
\right)^{2}}{e_{\beta~ in}^{8/5}}~I\left(e_{\beta~ in}, 0 \right)~,
\nonumber \\
K &\equiv& 0 ~.
\end{eqnarray}

Let us consider two particles characterized by the values $\beta_{1}$
and $\beta_{2}$. Moreover, let the particles have the same value of
$\bar{Q'}_{pr}$. If we are interested in times of stay of the particles
within an interval of semimajor axes ($a_{lower}$, $a_{upper}$), then
Eq. (77) yields: $ \tau_{1} / \tau_{2}$ $=$ $\beta_{2} / \beta_{1}$,
if the initial values of semimajor axes and eccentricities are equal
for both particles. On the basis of Eqs. (72)-(73) ($K$ $\equiv$ 0),
this result can
be approximately generalized also to the case of real solar
wind effect, under the assumption $\beta_{1}$, $\beta_{2}$ $\ll$ 1.

As for the secular evolution, under the assumption that particle's radius
does not decrease, we have to solve only one differential equation
Eq. (76); Eq. (75) immediately yields semi-latus rectum and one can easily
obtain semimajor axis
$a_{\beta}$ $=$ $p_{\beta}$ $/$ ($1- e_{\beta}^{2}$).

\subsection{Summary}
Our analytical approach shows that decrease of particle radius due to the solar wind
abrasion (corpuscular sputtering) generates increase of particle's semimajor axis
and eccentricity. This result is consistent with detailed numerical calculations
presented by Kocifaj and Kla\v{c}ka (2008). 

\section{Numerical results}
In this section we will concentrate on orbital evolution of interplanetary dust particle 
under the action of solar electromagnetic and corpuscular radiation when solar wind erosion is neglected. The results based on our new approach 
presented in Secs. 2-4
will be compared with the standard approach when only radial solar wind 
with constant $\eta$ is taken into account.

\subsection{Radial solar wind} 

\begin{figure}[h]
\begin{center}
\begin{minipage}{6cm}
\begin{center}
\includegraphics[height=0.22\textheight]{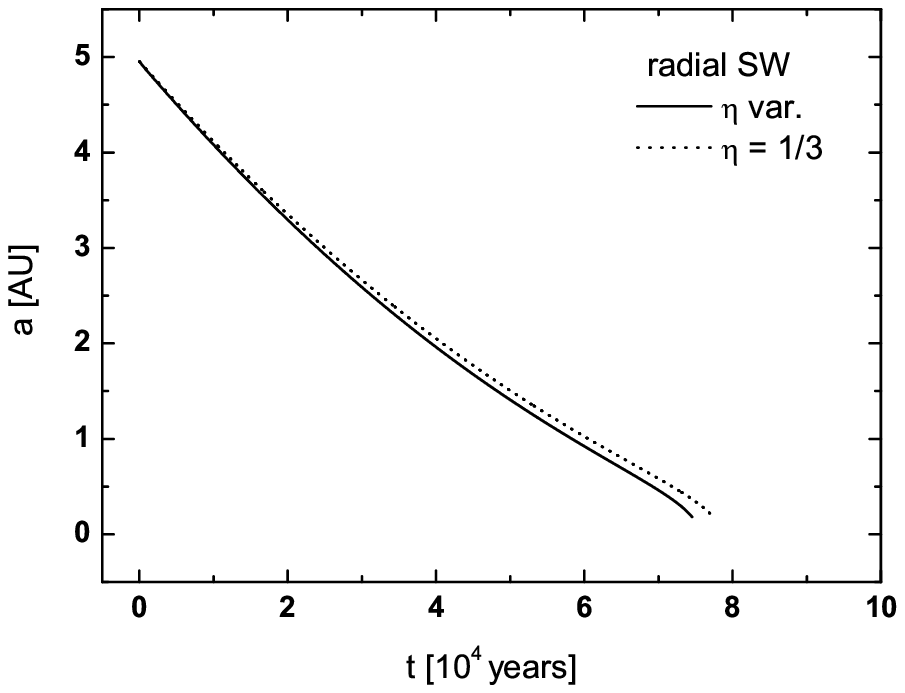}
\end{center}
\end{minipage}
\begin{minipage}{6cm}
\begin{center}
\includegraphics[height=0.22\textheight]{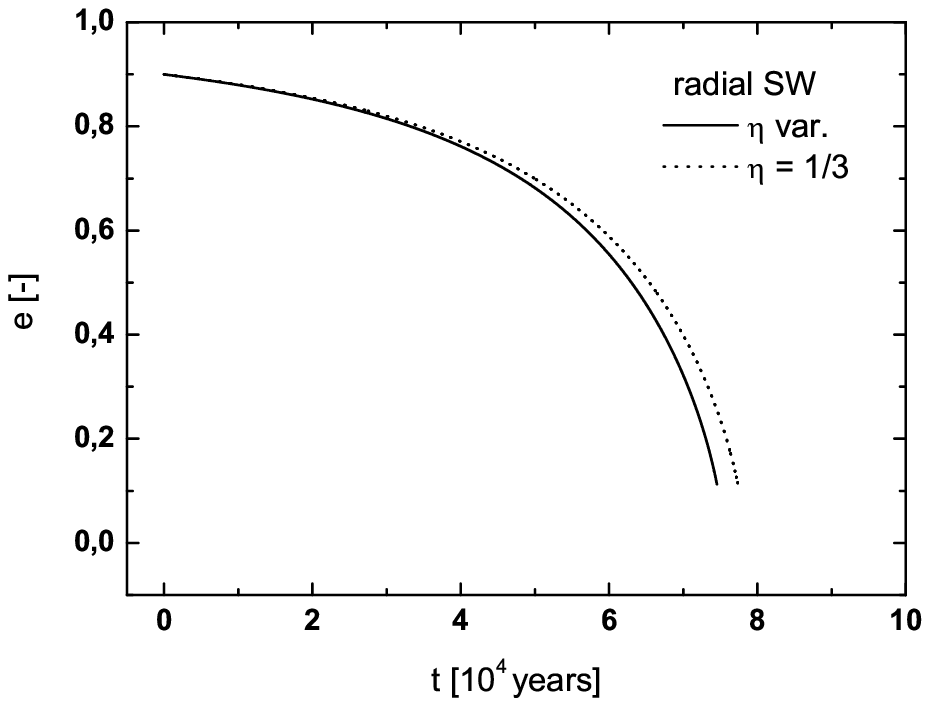}
\end{center}
\end{minipage}
\begin{minipage}{6cm}
\vspace{0.3cm}
\begin{center}
\includegraphics[height=0.22\textheight]{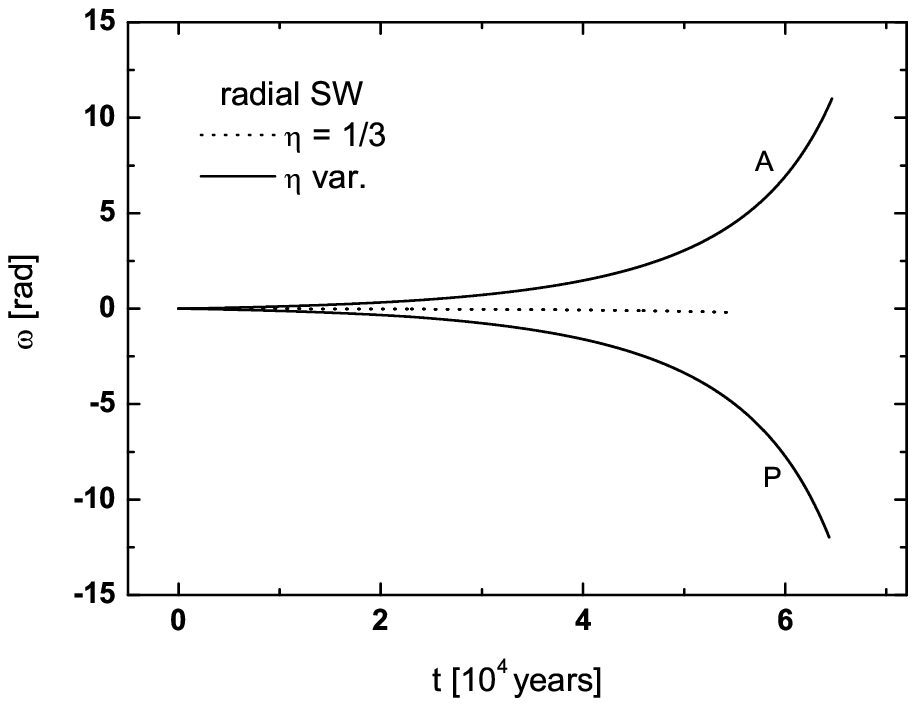}
\end{center}
\end{minipage}
\end{center}
\caption{Orbital evolution of spherical dust particle 
($\bar{Q}'_{pr}$ $=$ 1, $\beta =$ 0.01) 
for radial solar wind. Results for variable $\eta$ and standard value $\eta =$ 1/3 are compared.
Secular evolution of semimajor axis and eccentricity and also shift 
of perihelion are depicted. More realistic case yields that the
evolution of the shift of perihelion depends on the position 
(A - aphelion, P - perihelion) 
of the parent body at the time of the particle's ejection.
Effect of solar wind erosion is of negligible importance.}
\label{F1}
\end{figure}

Real flux density of solar wind energy and the approximation of a constant
flux ($\eta$ $=$ 1/3) are compared, for radial solar wind, in Fig. 1.
Eqs. (50) and (58) were numerically calculated for a
particle of fixed radius: $\beta$ $=$ 0.01, $\bar{Q}'_{pr}$ $=$ 1. 
 The orbital elements were calculated
for the central acceleration $-$ ($G M_{\odot} /r^{2}$) $\vec{e}_{R}$.  

The variable solar wind properties 
cause that semimajor axis exhibits a little faster decrease than the
standard approach when $\eta =$ 1/3 is used. The same holds also for
eccentricity of the particle. The time of spiralling toward the Sun
for variable $\eta$ is about 10 \% smaller than for the case of $\eta =$ 1/3.
The most significant difference between variable and constant $\eta$
exists for secular evolution of argument of perihelion. While the constant
$\eta$ produces no shift of perihelion, the more realistic approach
of variable $\eta$ produces nonnegligible shift of perihelion. Moreover,
the shift of perihelion depends on the initial position of the particle
in its orbit. Fig. 1 shows evolution of the shifts of perihelia for
two cases, for aphelion and perihelion ejections of the particle from 
a parent body (zero ejection velocity). Semimajor axis and eccentricity 
of the parent body were $a_{P}$ $=$ 5 AU, $e_{P}$ $=$ 0.9.   

\subsection{Real solar wind}

\begin{figure}[h]
\begin{center}
\begin{minipage}{6cm}
\begin{center}
\includegraphics[height=0.22\textheight]{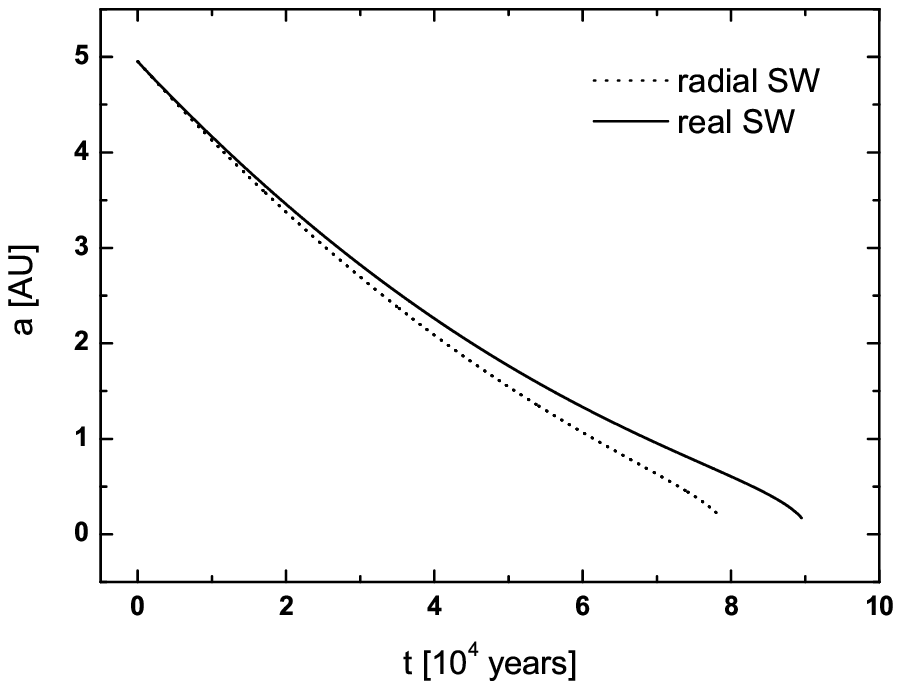}
\end{center}
\end{minipage}
\begin{minipage}{6cm}
\begin{center}
\includegraphics[height=0.22\textheight]{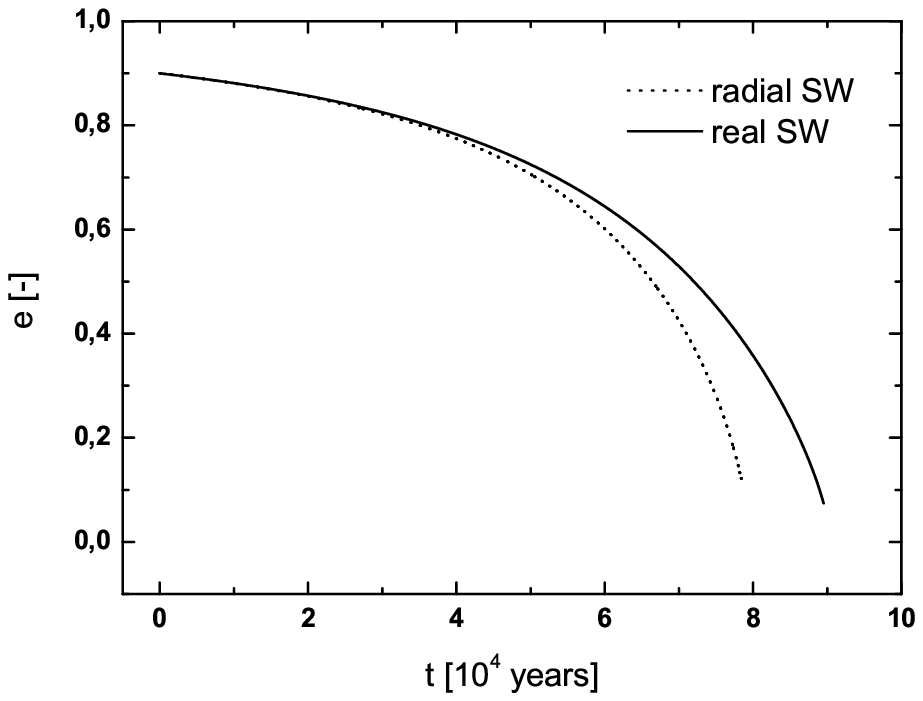}
\end{center}
\end{minipage}
\begin{minipage}{6cm}
\vspace{0.3cm}
\begin{center}
\includegraphics[height=0.22\textheight]{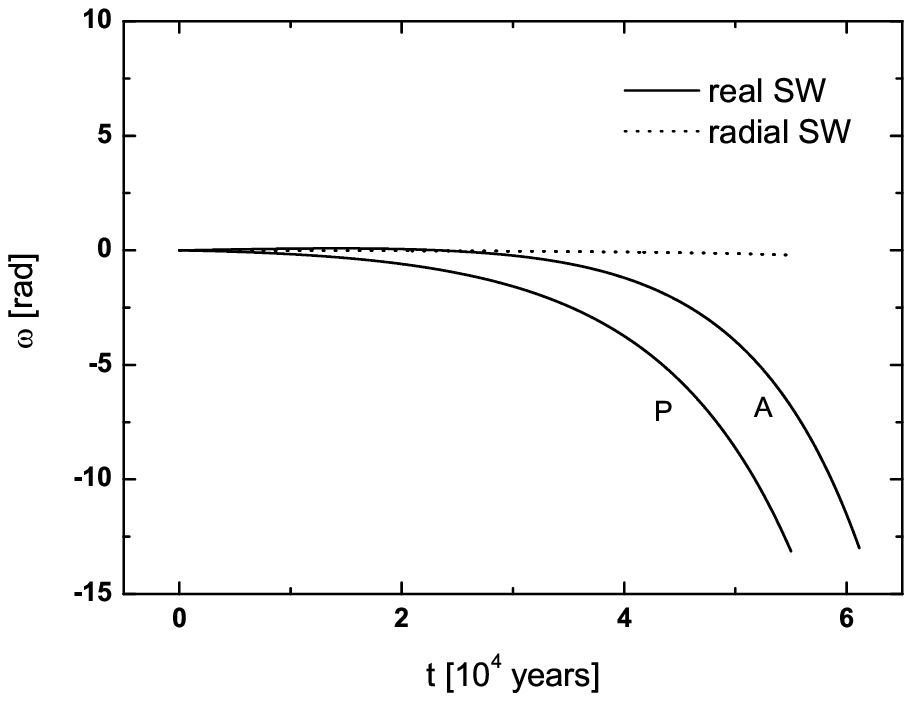}
\end{center}
\end{minipage}
\end{center}
\caption{Orbital evolution of spherical dust particle, 
 $\bar{Q}'_{pr}$ $=$ 1, $\beta =$ 0.01. 
Secular evolution of semimajor axis and eccentricity and also
shift of perihelion are depicted. 
Results for real solar wind (variable $\eta$ and non-radial component of solar wind velocity vector 
$\varepsilon =$ 2.9$^{o}$ are included) and radial solar wind (standard value $\eta =$ 1/3) are compared.
More realistic case yields that the evolution of the shift of perihelion 
depends on the position (A - aphelion, P - perihelion) 
of the parent body at the time of the particle's ejection.
Effect of solar wind erosion is of negligible importance.}
\label{F2}
\end{figure}

If we take into account even more realistic description of the
solar wind, when its nonradial velocity component is considered,
the resulting orbital evolution differs from the cases discussed in
Sec. 6.1. Fig. 2 compares the standard approach (radial solar wind,
$\eta$ $=$ 1/3) with the variable $\eta$ and $\gamma_{T} \ne$ 0.
Again, Eqs. (50) and (58) were numerically calculated for the
particle of fixed radius: $\beta$ $=$ 0.01, $\bar{Q}'_{pr}$ $=$ 1.
The orbital elements were calculated
for the central acceleration $-$ ($G M_{\odot} /r^{2}$) $\vec{e}_{R}$.  

The real solar wind causes that semimajor axis exhibits a little slower 
decrease than the standard approach (radial wind, $\eta =$ 1/3) is used. 
The same holds also for
eccentricity of the particle. The time of spiralling toward the Sun
for real solar wind action is about 15 \% greater than for the case of 
$\eta =$ 1/3 (one must be aware that this value holds for the case 
presented in Fig. 2 -- the real percentage may be greater/less for 
larger/smaller initial values of semimajor axis than for the initial 5 AU).   
The most significant difference between the usage of the real solar wind 
and the standard approach exists for secular evolution of argument of perihelion. While the constant $\eta$ produces no shift of perihelion, 
the realistic approach produces nonnegligible shift of perihelion. 
Moreover, the shift of perihelion depends on the initial position of the particle in its orbit. Fig. 2 shows evolution of the shifts of perihelia 
for two cases, the aphelion and perihelion ejections of the particle from 
a parent body (zero ejection velocity). Semimajor axis and eccentricity 
of the parent body were $a_{P}$ $=$ 5 AU, $e_{P}$ $=$ 0.9. 

\section{Discussion}
We have derived relativistically covariant equation of motion
for the action of solar wind corpuscles on motion of interplanetary
dust particles. As for spherical shape of the particles, the equation
of motion is represented by Eq. (29). It differs from the force
conventionally presented in literature (although only to the first
order in $\vec{v} / u$): \\
$\vec{F}_{sw}$ $=$ $F'_{sw}$ 
[ (1 $-$ 2 $\dot{r} / u$) $\vec{e}_{R}$ $-$
($r \dot{\theta} / u$)  $\vec{e}_{T}$ ] ,\\
where $\vec{v}$ is the velocity of the grain $\vec{v}$ $=$
$\dot{r}$ $\vec{e}_{R}$ $+$ $r \dot{\theta}$ $\vec{e}_{T}$, 
$u$ is the heliocentric solar-wind speed and $F'_{sw}$ is the 
force on the dust for $\vec{v}$ $=$ 0 (Minato {\it et al.} 
2004 -- Sec. 2.1; equivalent force is presented in Eq. (7.10) by 
Mann 2009; see also p. 12 in Burns {\it et al.} 1979).

We have to stress that the standard form corresponds to Eq. (32)
if $x'$ $=$ 1 (reality is: 1 $< x' <$ 3, approximately) and not to
physically correct form given by Eq. (29). Eq. (29) yields the
correct limiting result $u \rightarrow c$ equivalent to the P-R effect.

As for practical application of the above discussed physical results,
Eqs. (56)-(59) are astronomically relevant.
Decrease of mass of the particle may cause its spiralling outward
from the Sun and not toward the Sun, as it is commonly accepted for the
action of solar wind on interplanetary dust particles. This result is
evident from analytical equations presented in Sec. 5 (see, e. g.,
Eqs. 61-62) and detailed numerical calculations confirming this result
can be found in Kocifaj a Kla\v{c}ka (2008). Sec. 6 concentrates on
the action of solar wind on motion of interplanetary dust particles
for the cases when variable flux of solar wind energy and non-radial
solar wind velocity are considered. The most evident difference between
the standard and more realistic approaches is represented by the shift
of perihelion (secular evolution of argument of perihelion). It is
generally believed that the shift of perihelion does not exist, both
for the solar wind effect and the P-R effect. However, the real action of
the solar wind differs from the action of the P-R effect -- the P-R effect
really produces no shift of perihelion. The real shift of perihelion
depends on the initial orbital position of the particle. The non-radial
solar wind velocity can lead to outspiralling from the Sun, in the
region of outer planets. The particle may or may not spiral toward 
the Sun due to the simultaneous action of the P-R effect and solar 
wind effect (see also Kla\v{c}ka {\it et al.} 2008).
 
\section{Conclusion}
Relativistically covariant equation of motion for
arbitrarily shaped dust particle under the action of
solar wind is derived. Change of the particle's mass is an
indispensable part of the space-time formulation of the equation of
motion for the action of the solar wind. The solar wind effect would
reduce to the Poynting-Robertson effect, in the limiting case when:
i) solar wind speed would tend to the speed of light, ii) no decrease of
mass of the interplanetary dust particle would exist, and,
iii) the velocity of the solar wind would be radial.
However, the solar wind may have qualitatively different effect on orbital
evolution of interplanetary dust particle, since the points ii) and iii)
are not fulfilled, in general. The decrease of mass of the interplanetary 
dust particle and non-radial component of the solar wind velocity 
may cause outspiralling of the particle from the Sun. Time variable solar 
wind leads to the shift of perihelion of the particle. The found results
may have important consequences for evolution of dust disks in the 
vicinity of stars with stellar winds.

\appendix
\section{Emission from the particle}
The other possible force influencing dynamics of dust particle 
may originate from an emission, e.g., radioactive decay. 
Let the particle emits an energy $E'_{em}$ per unit time due to 
the emission in the proper reference frame. We will suppose that 
this emission is represented by the flux of particles with a speed 
$u'_{em}$. Further, we declare the orthonormal vector basis 
$\{\vec{f'}_{j};~j=1,~2,~3\}$, as it was used in Sec. 2.2. The 
corresponding velocities are: 
$\vec{u'}_{em,j}~=~u'_{em} \vec{f'}_{j},~ j=1,~2,~3$.  

The outgoing four-momentum of the emission per unit time, 
in the proper reference frame of the particle, is
\begin{equation}\label{A1}
p'^{\mu}_{em}~=~\left(\frac{1}{c}~E'_{em}~;~\frac{1}{c}~E'_{em}~
\sum_{j=1}^{3}~r'_{j}~\frac{\vec{u'}_{em, j}}{c}\right)~,  
\end{equation}
where $r'_{j}$ ($j=1,~2,~3$) are dimensionless coefficients expressing 
the part of the total flux of radiation which is emitted in the 
corresponding directions.  

Lorentz transformations of (A.1) yield the outgoing four-momentum 
per unit time in stationary reference frame: 
\begin{equation}\label{A2}
p^{\mu}_{em}~=~\frac{1}{c}~E'_{em}~\frac{U^{\mu}}{c}~+~\frac{1}{c}~E'_{em}~
\sum_{j=1}^{3}~r'_{j} \left(\xi^{\mu}_{em,j}~-~\frac{U^{\mu}}{c}\right)~,
\end{equation}
where 
\begin{eqnarray}\label{A3}
\xi^{\mu}_{em,j} &=& \left(\frac{1}{\omega_{em,j}}~;~\frac{1}{\omega_{em,j}}~
\frac{\vec{u}_{em,j}}{c}\right)~,
\nonumber\\
\omega_{em,j} &\equiv& \gamma\left(v\right) \left(1~-~
\frac{\vec{v}\cdot\vec{u}_{em,j}}{c^{2}}\right)~,
\nonumber\\
\vec{u}_{em,j} &=& \left[\gamma\left(v\right)\left(1+
\frac{\vec{v}\cdot\vec{u'}_{em,j}}{c^{2}}\right)\right]^{-1} 
\left\{\vec{u'}_{em,j}+\left[\left(\gamma\left(v\right)-1\right) 
\frac{\vec{v}\cdot\vec{u'}_{em,j}}{\vec{v}^{2}}+\gamma\left(v\right)\right] 
\vec{v}\right\}~,
\nonumber\\
j &=& 1,~2,~3~.
\end{eqnarray}
Thus, expression on the right-hand side of Eq. (A.2) should be added to 
the right-hand side of Eq. (21) when we would like to take into account 
also the effect of the emission of dust particle.  

\section*{Acknowledgement}
The authors want to thank to Prof. \v{S}ev\v{c}ovi\v{c} for discussion
on solution of Burgers equation. This work was partially supported by the Scientific Grant Agency VEGA, Slovakia,
grant No. 2/0016/09 and by the Comenius University grant UK/405/2009.

\end{document}